\documentclass[aip,amsmath,amssymb,reprint]{revtex4-1}

\usepackage{graphicx}
\usepackage{dcolumn}% Align table columns on decimal point
\usepackage{bm}% bold math\usepackage{amsfonts}
\usepackage{color}
\usepackage[utf8]{inputenc}
\usepackage[T1]{fontenc}

\usepackage[usenames,dvipsnames]{xcolor}
\usepackage[normalem]{ulem}   %% e.g. fuer \sout{} zum Durchstreichen von text!!

\usepackage{cancel}

\newcommand\redsout{\bgroup\markoverwith
{\textcolor{red}{\rule[+0.45ex]{2pt}{1.2pt}}}\ULon}

\usepackage{soul} %% e.g. for highlighting of text with \hl{}

\newcommand{\Eq}[1]{Eq.~\eqref{#1}}
\newcommand{\Eqs}[1]{Eqs.~\eqref{#1}}
\newcommand{\ui}{\text{i}}

\newcommand{\iexp}{\text{i}}

\newcommand{\la}{\langle}
\newcommand{\ra}{\rangle}
\newcommand{\df}{\delta\!f}

\newcommand{\deltath}{\delta_{\rm th}}
\newcommand{\no}{\nonumber\\}

\newcommand{\enl}{\tilde \epsilon}
\newcommand{\znl}{\tilde \zeta}

\begin{document}

%%%%%%%%%%%%%%%%%%%%%%%%%%%%%%%%%%%%%%%%%%%%%%%%%%%%%%%%%%%%%%%%%%%%
%%%%%%%%%%%%%%%%%%%%%%%%%%%%%%%%%%%%%%%%%%%%%%%%%%%%%%%%%%%%%%%%%%%%
\title{Parametric Effects in Circuit Quantum Electrodynamics (Review Article)}
\author{Waltraut Wustmann}
\affiliation{Laboratory for Physical Sciences, College Park, MD 20740, USA}
\author{Vitaly Shumeiko}
\affiliation{Department of Microelectronics and Nanoscience, Chalmers University of Technology, 41296 G\"oteborg, Sweden}


%\date{27 February 2019}

\begin{abstract}
We review recent advances in the research on quantum parametric phenomena in superconducting circuits with Josephson junctions. We discuss physical processes in parametrically driven tunable cavity and outline theoretical foundations for their description.  Amplification and frequency conversion are discussed in detail for degenerate and non-degenerate parametric resonance, including quantum noise squeezing and photon entanglement. Experimental advances in this area played decisive role in successful development of quantum limited parametric amplifiers for superconducting quantum information technology. We also discuss nonlinear down-conversion processes and experiments on self-sustained parametric and subharmonic oscillations.  
\end{abstract}

\maketitle
%%%%%%%%%%%%%%%%%%%%%%%%%%%
%%%%%%%%%%%%%%%%%%%%%%%%%%%%

\section{Introduction}

In this contribution to special issue of Low Temperature Physics journal commemorating 100 year anniversary of B.I. Verkin we survey the progress in the research on quantum parametric phenomena in superconducting electrical circuits.
In his leadership role as a director of large research institution B.I. Verkin gave preference to new practical developments, but at the same time he paid great attention to fundamental research.
This fruitful combination of the fundamental and applied well describes the subject of this article.
Exploration of quantum parametric phenomena is the part of recently emerging and rapidly growing field of circuit quantum electrodynamics (c-QED) -  a quantum information technology based on superconducting Josephson junctions. A typical experimental c-QED device, see e.g. \cite{ChowNatC2014,DiCarloNatC2017}, contains a network of nonlinear oscillators - Josephson junctions, and linear oscillators - high quality superconducting resonators. The network operates in the quantum regime at frequencies of few GHz and temperature of tens miliKelvin.  For the reviews on the Josephson junction based quantum bits and c-QED see Refs.\cite{MakhlinRMP2001,MartinisArxiv2004,WendinLTP2007,SchoelkopfNat2008,KockumPR2017,WendinRPP2017} and references therein.

Recent interest to quantum parametric effects in a microwave domain was motivated by practical need to amplify extremely weak, of single-photon intensity, microwave signals carrying an information about qubit states. Required noise performance of amplifiers is therefore demanding, it must be close to the limit imposed by the Heisenberg uncertainty principle.  During last decade a great effort was made to develop quantum limited parametric amplifiers   \cite{BergealNature2010,EichlerPRL2011,BergealPRL2012,RocETAL2012,FluETAL2012,EichlerPRL2012,YamETAL2008,VijayPRL2011,RistePRL2012,MenzelPRL2012,NakamuraAPL2013,VionPRB2014,DeppeNJP2015}. Success of this work was an important step in advancing research on superconducting qubits and development of c-QED technology.
 
The most of known natural and engineered parametric phenomena in mechanics, hydrodynamics, plasma physics, etc., occur under classical physics conditions. In the 
c-QED, similar to the quantum optics, quantum properties of electromagnetic field - the photon statistics and correlations -
 come to the first place.  Due to the new parameter regimes available in the quantum microwave optics due to strongly nonlinear properties of the Josephson junctions, a number of  phenomena, in principle known in theory, become available in the experiment. Notable examples are
 an ultrastrong light-matter interaction \cite{GrossNatP2010,LupascuNatP2017,SembaNatP2017},  dynamical Casimir effect\cite{WilsonNature2011}, multiphoton quantum cat states \cite{VistakisSci2013,LeghtasSci2015}.

In this article we review some theoretical results and experimental observations on parametric effects in tunable superconducting resonators. Superconducting resonators, being essentially linear electromagnetic devices acquire nonlinear property due to the coupling to the Josephson junctions. Nonlinear resonators were employed for parametric amplification,  frequency conversion, demonstration of noise squeezing and photon entanglement. A tunable cavity belongs to this family of  resonators. Here a dc SQUID is attached to the resonator, which allows controlling the frequency of the resonator  by varying the magnetic flux through the SQUID \cite{Wallquist2006,Sandberg2008}. Rapid temporal modulation of the magnetic flux allows one to achieve the amplification effect\cite{YamETAL2008}, and excite the parametric resonant oscillation\cite{WilsonPRL2010}. The method of the flux pumping in tunable cavity can be compared to the optomechanics, where motion of mirrors of optical resonators produces the parametric effect\cite{AspelmeyerRMP2014}. Spectacular manifestation of this effect is the dynamical Casimir effect (DCE) - the quantum effect of creation of  photons from the vacuum by moving mirrors\cite{Moore1970}. Analogy between the DCE and parametric effect in tunable cavity led to prediction\cite{JohanssonPRL2009,JohanssonPRA2010} and  observation\cite{WilsonNature2011} of the DCE with flux pumped SQUID. 

Alternative method of parametric excitation, the  current pumping was used in several of cited experiments. With this method, which is similar to the nonlinear optics, a strong signal, current pump, is injected into the resonator to stimulate nonlinear intermode interaction. 

The physics of the tunable cavity is at the border of two physics areas - nonlinear mechanics and nonlinear optics. Dynamics of the field of the cavity modes constitutes the mechanical aspect, which makes relevant all the accumulated knowledge about parametric resonance in nonlinear classical oscillators \cite{NayfehBook,BogoliubovBook,JordanBook}. On the other hand, inelastic scattering of quantum electromagnetic field by parametrically driven cavity is the optical aspect, that includes the amplification effect, frequency conversion, and generation of a nonclassical microwave field. 

The paper has the following structure. In Sec.~\ref{sec:TunableCavity} we describe a theoretical model of the tunable cavity, which is based on the method of quantum Langevin equation, and introduce the resonance approximation to describe the non-degenerate and degenerate parametric resonance. In Sec.~\ref{sec:Response} we describe the cavity linear and nonlinear response in the regime of amplification and frequency conversion, and then,  in Sec.~\ref{sec:oscillation}, proceed to the discussion of self-sustained parametric and subharmonic oscillations. In Sec.~\ref{sec:quantum}  we turn to the quantum properties of microwave field generated by tunable cavity, we discuss the quantum squeezing and entanglement, and analyze the efficiency of signal amplification in the terms of the signal to noise ratio.

%%%%%%%%%%%%%%%%%%%%%%%%%%%%%%%%%%%%
\section{Description of the device}
\label{sec:TunableCavity}
\begin{figure}[h]
\centering
 \includegraphics[width=0.7\columnwidth]{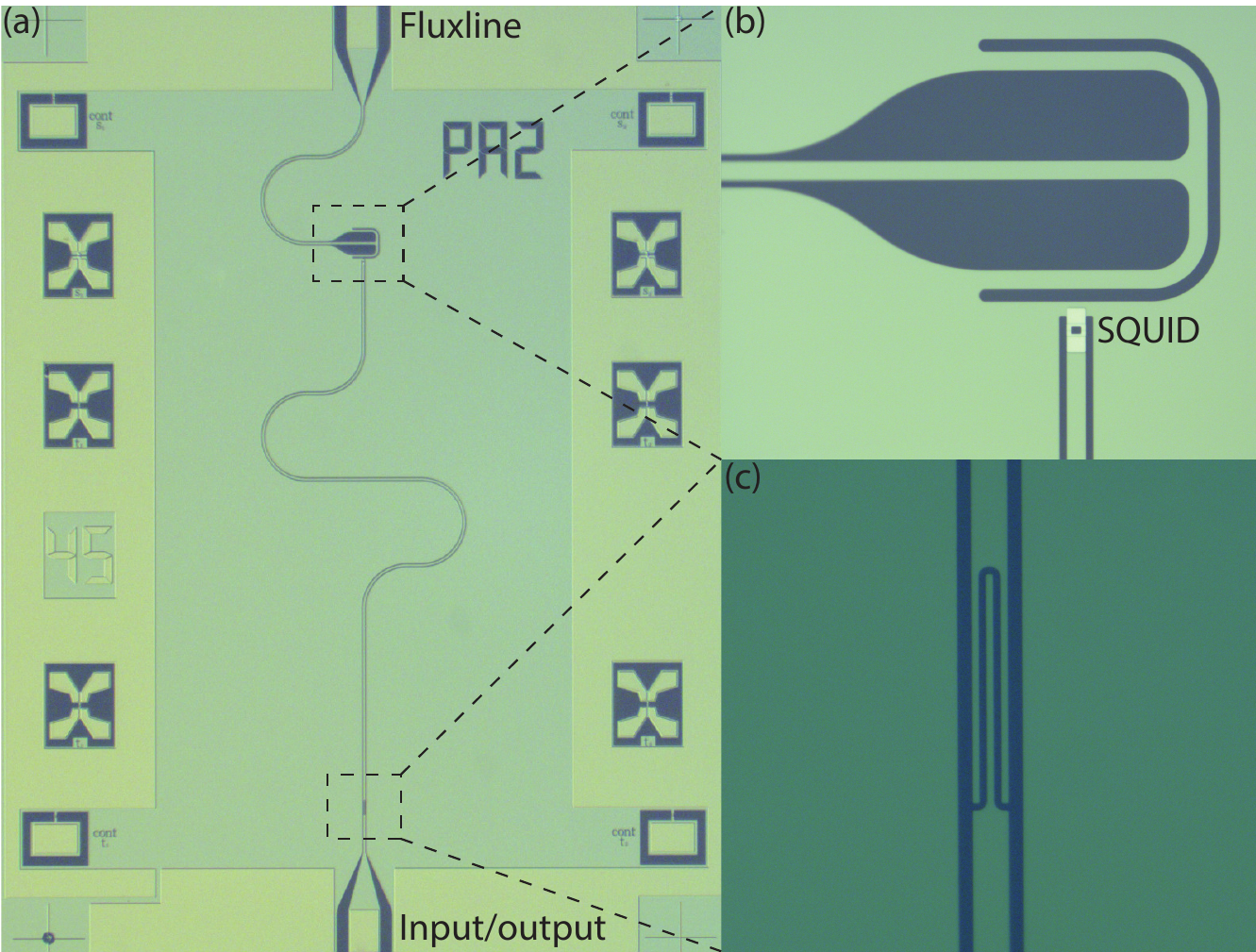}
\caption{Micrographs of a typical tunable resonator: (a) full chip (5x7 mm) with the  
flux line at the top, and the input/output port with the coupling capacitor at the bottom;  (b) SQUID together with the inductive coupling between the flux line and the SQUID loop; (c) Coupling capacitor. (Adopted from\cite{AndreasLic2017}, courtesy of A. Bengtsson.)  }
\label{Fig:CavityDevice}
\end{figure}

Tunable cavity  is a $\lambda/4$ superconducting resonator, made with a segment of a coplanar waveguide, which is galvanically connected to a dc SQUID at one end, and to a transmission line at the other end\cite{Sandberg2008,YamETAL2008}, see Fig.~\ref{Fig:CavityDevice}. 
The plasma frequency of the SQUID is much larger than the resonator frequency thus the former acts as a tunable nonlinear inductance. The presence of the SQUID makes the eigenfrequency spectrum of the resonator non-equidistant    and also introduces a nonlinearity. Typical device operates in a quantum regime at frequencies $\omega$ from few to tens GHz, at temperature $T\sim 20$ mK, has  large quality factor, $Q\sim 10^{4}$, dominated by external losses,  and small nonlinearity, the Kerr coefficient $\sim 10^{-5}\omega$.

\begin{figure}[h]
\centering
 \includegraphics[width=0.8\columnwidth]{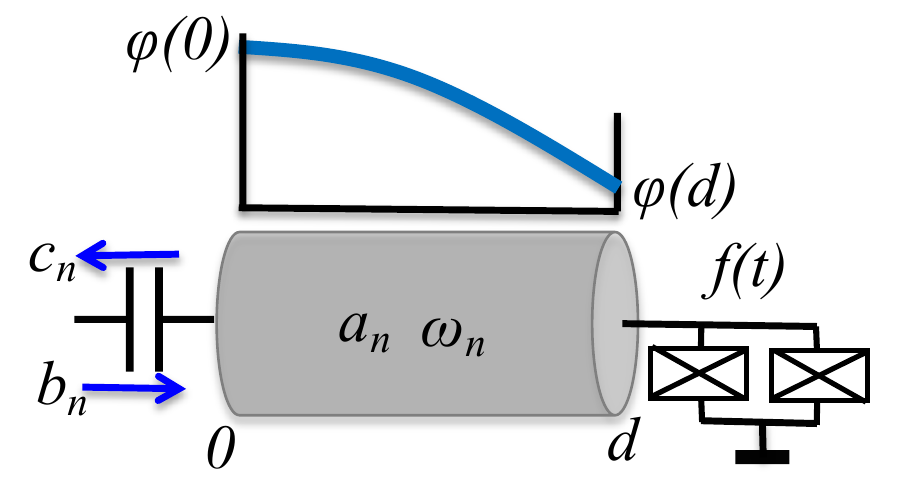}
\caption{Sketch of tunable cavity: $a_n(t)$ is the complex amplitude of $n$-th cavity eigen mode, $b_n(t)$ and $c_n(t)$ are the input and output field amplitudes, respectively; the cavity eigen frequencies $\omega_n$  are controlled by magnetic flux $f(t) = F + \df  \cos\Omega t$; the plot above the cavity illustrates spatial field distribution of the fundamental mode, $\phi_0(x,t)$.}
\label{Fig:TunableCavity}
\end{figure}

A spatial profile of the field in the resonator is illustrated in  Fig.~\ref{Fig:TunableCavity}. The magnitude of the field at the resonator end connected to the SQUID depends on the SQUID inductance and can be controlled by varying magnetic flux applied to the SQUID. This results in variation of the cavity eigenfrequency spectrum\cite{Wallquist2006}. Rapid temporal modulation of the magnetic flux produces the parametric effect\cite{WustmannPRB2013}. 

Theoretical description of physical processes in tunable cavity is based on the Lagrangian description of electrical circuits \cite{YurkeDen1984,Devoret2004,WendinLTP2007,HasslerPRB2016}. The dynamical variable here is a 1D field, $\phi(x,t)$, that describes the spatial distribution of the superconducting phase along the cavity. The corresponding Lagrangian has the form\cite{Wallquist2006,WustmannPRB2013},
\begin{eqnarray}\label{Lagrangian_start}
{\mathcal L} = \left({\hbar \over 2e}\right)^2 {C_0\over 2} \int_0^d dx \,( \dot\phi^2 - v^2 \phi'^2) + 2 E_J\cos {f(t)\over2} \cos \phi(d) , \nonumber
\end{eqnarray}
and contains two parts, the cavity part represented by the integral term, and the SQUID inductance part (small SQUID capacitance plays a minor role and is omitted here). In this equation, $v=1/\sqrt{C_0 L_0}$ is the electromagnetic wave velocity,  $C_0$ and $L_0$ are the cavity specific capacitance and inductance, respectively, $E_J$ is the Josephson energy of a single junction, and $f = F + \df(t)$ is the magnetic flux threading the SQUID (in units of 
$\hbar/2e$); it consists of a constant bias, $F$, and  a temporal flux modulation, $\df(t) \ll 1$. 

Variation over variables $\phi(x,t)$, $\phi(d,t)$, and $\phi(0,t)$ yields the linear wave equation, 
$\ddot\phi(x,t) - v^2\phi''(x,t) = 0$, and a nonlinear boundary condition, $ \gamma d\phi'(d,t) +  \sin\phi(d,t) = 0$, where $\gamma = E_{L,cav} / 2 E_J \cos(F/2) \ll 1$ is a participation ratio of the SQUID vs cavity inductances. 
The second boundary condition, $\phi'(0) = 0$, defines the spatial profiles of the eigen modes, $\phi(x) \propto \cos kx$. The latter equation together with linearized boundary condition yield the cavity spectral equation,
\begin{eqnarray}
k_n d \tan k_nd  = {1\over \gamma} \gg 1, \quad \omega_n = vk_n\,.
\end{eqnarray}

To derive the cavity quantum Hamiltonian we expand the cavity field over eigenmodes,
\begin{eqnarray}\label{eq:modeexpansion}
\phi(x,t) &=& \sqrt{4\pi Z_0\over R_k} \sum_{n=0}^\infty {\cos k_n x \over \sqrt{k_nd}} \,(a_n(t) + a_n^\ast(t)), 
\end{eqnarray}
where $a_n(t)$ is the eigenmode complex amplitude,  $Z_0 = \sqrt{L_0/C_0}$ is the cavity impedance, and $R_k= h/2e^2$ is the quantum resistance,  substitute this expansion into the Lagrangian, and after some algebra arrive at the classical Hamiltonian \cite{WustmannPRB2013}, 
\begin{eqnarray}\label{eq:Hcav}
&& H = \sum_n \hbar \omega_n a_n^\ast a_n  + V[\phi] \,, \\
\label{eq:V}
&& V[\phi]  = - 2E_J \left(\cos{f(t)\over 2} \cos\phi(d,t) + \cos{F\over 2} {\phi^2(d,t)\over 2}\right).     
\end{eqnarray}
Quantum version of this Hamiltonian is obtained by imposing the bosonic commutation relations  on the mode amplitudes, $[a_n(t)\,,\, a_m^\dag(t)\,] = \delta_{nm}$.

%%%%%%%%%%%%%%%%%%%%%
\subsection{Langevin equation}
Equation (\ref{eq:Hcav}) describes the dynamics of the closed cavity disconnected from the environment.  Capacitive 
coupling of the cavity to the transmission line allows one to probe the cavity internal state and also to explore the cavity response to driving electromagnetic signals.  At the same time, this exposes the cavity to an environmental noise that leads to cavity damping. 

A suitable way to describe the dynamics of open cavity is to formulate the Langevin equation for the mode Heisenberg operators \cite{GarZol2000}, which has the form in the present case\cite{WustmannPRB2013}, 
\begin{eqnarray}\label{eq:langevin}
i \dot a_n - \omega_n a_n - {1\over \hbar} [\,a_n \, , V[\phi(d)]\,]   + i\Gamma_n a_n
= \sqrt{2\Gamma_{n0}} \,b(t)
\,.
\end{eqnarray}
Here $b(t)$ refers to incidental external field expressed through environmental electromagnetic modes, $a_k(t)$, \cite{ColGar1984,Yurke_DruFic2004} 
\begin{eqnarray}
b(t) = \sqrt{v\over 2\pi } \int _{0}^{\infty} dk \;a_k(t_0)e^{-i\omega_k(t-t_0)} \,,
\end{eqnarray}
that may also include the probing tones.The operators $b(t)$ satisfy the commutation relation, $[ b(t)\,,\, b^\dag(t')] = \delta(t-t')$. The rate
\begin{equation}\label{eq:Gn0}
 \Gamma_{n0} = \omega_n \left( \frac{C_c}{C_0d} \right)^2 k_n d \,, 
\end{equation}
quantifies external losses due to coupling to the transmission line  through  capacitance $C_c$. 
 $\Gamma_n$ in \Eq{eq:langevin} refers to the total  losses, which  include both external and internal losses. In what follows we will neglect the latter and suppress index $0$ in $\Gamma_{n0}$.

Full description of the open cavity is completed with an  equation for output field quantified with operator $c_n(t)$.  This equation has the form of an input-output relation \cite{ColGar1984},
\begin{equation}\label{eq:cb_general}
c_n(t) = b(t) - i\sqrt{2\Gamma_{n0}} \,a_n(t) 
 \,.
\end{equation}
%

%%%%%%%%%%%%%%%%%%%%
\subsection{Resonance approximation}
\label{sec:Resonance}

To study complex nonlinear equations like  \Eq{eq:langevin} one needs to resort to some simplifying assumptions. 
A usual assumption refers to small value of the field amplitude $\phi(d,t)\ll1$. This allows for series expansion of the cosine function in \Eq{eq:V}, and keeping the lowest relevant nonlinear terms. Another simplified assumption concerns small amplitude of the flux modulation, $\delta f(t) \ll 1$. Under this assumption one may linearize the potential $V$ in \Eq{eq:V} with respect to $\delta f(t)$. 

These simplifications, however, are not sufficient  because of the presence of resonances. The resonance, i.e. coincidence of driving frequency with some combination of system internal frequencies, strongly affects the system dynamics, which cannot be treated with simple perturbative methods \cite{NayfehBook,BogoliubovBook}. Even small nonlinearity and weak parametric drive produce a deviation from the linear behaviour, which is slow on the time scale of the linear oscillation and is large in amplitude. Formulation of equations describing such a secular resonant dynamics is the subject of the resonance approximation. 

%%%%%%%%%%%%%%%%%%%%%
\subsubsection{Nondegenerate parametric resonance}
Particular simplification of general Langevin equation \eqref{eq:langevin} depends on the resonance under consideration. We start with the 
situation when magnetic flux is harmonically modulated   with frequency close to 
the sum of two cavity modes, $\delta f(t) = \delta f \,\cos\Omega t$, $\Omega = \omega_n + \omega_m + 2\delta$, 
where $\delta \ll \omega_n$ is a small 
detuning from exact resonance. In the lowest order such a modulation drives the frequencies of  both modes leading 
to non-degenerate parametric resonance. One has to note that the excitation of only two selected cavity modes essentially relies on the non-equidistance of the cavity spectrum, which must exceed the mode bandwidth. 

The resonance dynamics is commonly described in 
the rotating frame,  $a_n(t) \rightarrow  e^{-\iexp(\omega_n+\delta) t}  a_n(t)$, taking advantage of slow time 
variation of the Heisenberg operators in this frame. Averaging \Eq{eq:Hcav} over fast rotations, we arrive at the 
Hamiltonian describing cavity resonant dynamics \cite{Wustmann2017},
\begin{eqnarray}\label{eq:H_ndeg}
 H &= & -  \sum_{j=n,m} \left[ \hbar \delta  a_j^\dag a_j 
 +  {\hbar \alpha_j\over2} \left(a_j^\dag a_j \right)^2 \right]  \no
  &-& 2\hbar \alpha_{nm} (a_n^\dag a_n a_m^\dag a_m) 
 - \hbar  \epsilon_{nm} \left( a_n a_m +  a_n^\dag a_m^\dag \right) 
 \,.
\end{eqnarray}
Here we have  retained the lowest order terms in the expansion of $\cos\phi$ in the Josephson potential in \Eq{eq:V}: the quadratic term in the part containing flux modulation, and the quartic terms in the static part. The former one parametrically couples the modes with the strength, 
\begin{eqnarray}\label{eq:def_epsilon}
  \hbar\epsilon_{nm}  =  \frac{\delta f}{2} E_J \sin{F\over 2} s_n s_m, \;\; s_j = 
\sqrt{4\pi Z_0\over R_k} \left(\frac{\cos k_j d}{\sqrt{ k_j d} }\right)
 \,, \no
\end{eqnarray}
while the latter ones describe the self-Kerr effect and the cross Kerr effect quantified with respective coefficients,
\begin{eqnarray}\label{eq:def_alpha_n}
 \hbar\alpha_j  = E_J\cos {F\over 2} \,s_j^4 
 \quad \alpha_{nm} = \sqrt{\alpha_n \alpha_m} 
 \,,
\end{eqnarray}
(in \Eq{eq:H_ndeg} we skipped small corrections, $\propto \alpha_j$,   to detuning $\delta$).
 
Hamiltonian (\ref{eq:H_ndeg}) is equivalent to the one for two nonlinear oscillators with parametrically driven coupling. It can be equivalently written in terms of quadratures (coordinate and momentum), 
$q_n = (a_n + a^\dag_n)/\sqrt 2$ and $p_n = -i(a_n - a^\dag_n)/ \sqrt 2$,
\begin{eqnarray}\label{eq:Heff_ndeg}
 && H/\hbar = -  \sum_{j=n,m} \left[ \frac{\delta}{2} (q_j^2+p_j^2)  
+  \frac{\alpha_j}{8 } \left(q_j^2 + p_j^2\right)^2\right] \\
&-& {\alpha_{nm}\over 2} (q_n^2+p_n^2)(q_m^2+p_m^2)
 - \epsilon_{nm} (q_nq_m - p_n p_m)\,. \nonumber
\end{eqnarray}

The corresponding Langevin equations consist of two coupled equations,
\begin{eqnarray}\label{eq:EOM_ndeg}
 \ui \dot a_n  &+&  \left[ \delta  + \ui \Gamma_n  + \alpha_n (a_n^\dag a_n) 
 + 2 \alpha_{nm} (a_m^\dag a_m )  \right] \,a_n + \epsilon_{nm} a_m^\dag  \no
 &=& \sqrt{ 2\Gamma_{n}} \,b_n(t) 
 \,,
\end{eqnarray}
with the input fields, $b_{n}(t)$, being written in the respective rotating frames. 
The input-output relations in \Eq{eq:cb_general} retain their form in the rotating frame.

The resonance approximation relies on the separation of the frequency of time  variation of amplitude $a_n(t)$ from the mode frequencies. This implies that all the coefficients in \Eq{eq:EOM_ndeg} respect the constraint,
\begin{eqnarray}\label{constraints2}
\delta, \; \alpha_j ,  \; \Gamma_j , \; \epsilon_{nm}  \ll \omega_n - \omega_m \sim \omega_j\,.
\end{eqnarray}
It is worth to mention that small values of the Kerr coefficients and the pumping coefficients rely on small values of parameters $s_j$ in \Eq{eq:def_epsilon}, which are provided by the small value of factors, 
$\cos k_jd \sim \gamma \ll 1$, in addition to the small ratio, $Z_0/R_k \ll 1$.

%%%%%%%%%%%%%%%%%%%%%%
\subsubsection{Degenerate parametric resonance}
There is a special, degenerate form of the parametric resonance, when only one mode is excited. In this case, $\omega_n = \omega_m$ and the flux is modulated with frequency close to twice the mode frequency,  
$\Omega = 2\omega_n + 2\delta$. This regime of the modulation of mode frequency 
 is mostly studied in literature. The Hamiltonian (\ref{eq:Hcav}) reduces in this regime to the form\cite{WustmannPRB2013}, 

\begin{eqnarray}\label{eq:Heff_deg}
 H = -  \hbar \delta  a_n^\dag a_n  -  {\hbar \alpha_n\over2} (a_n^\dag a_n)^2  
 - {\hbar  \epsilon_n\over 2} \left( a_n^2 +  a_n^{\dag 2}  \right) .
\end{eqnarray}
In  terms of quadratures, this Hamiltonian is equivalent to the one of parametrically driven Duffing oscillator \cite{DykmanPRE1998},
\begin{eqnarray}\label{eq:Heff_deg_pq}
 H/\hbar = - {\delta\over 2}  ( q_n^2 + p_n^2)  -  {\alpha_n\over8} \left(q_n^2 + p_n^2 \right)^2 
 - \epsilon_n \left( q_n^2 -  p_n^{2}  \right) . \no
\end{eqnarray}
The Langevin equations reduce to a single equation,
\begin{eqnarray}\label{eq:EOM_deg}
 \ui \dot a_n  +  \left[ \delta  + \ui \Gamma_n  + \alpha_n (a_n^\dag a_n) \right] \, a_n + \epsilon_n a_n^\dag  
 = \sqrt{ 2\Gamma_{n}} \,b_n(t) \,.
\end{eqnarray}
%

%%%%%%%%%%%%%%%%%%%%%%%%%%%%%%%%%%%
\subsection{More complex configurations}
\label{sec:2cavity}
\begin{figure}[h]
\centering
\includegraphics[width=0.7\columnwidth]{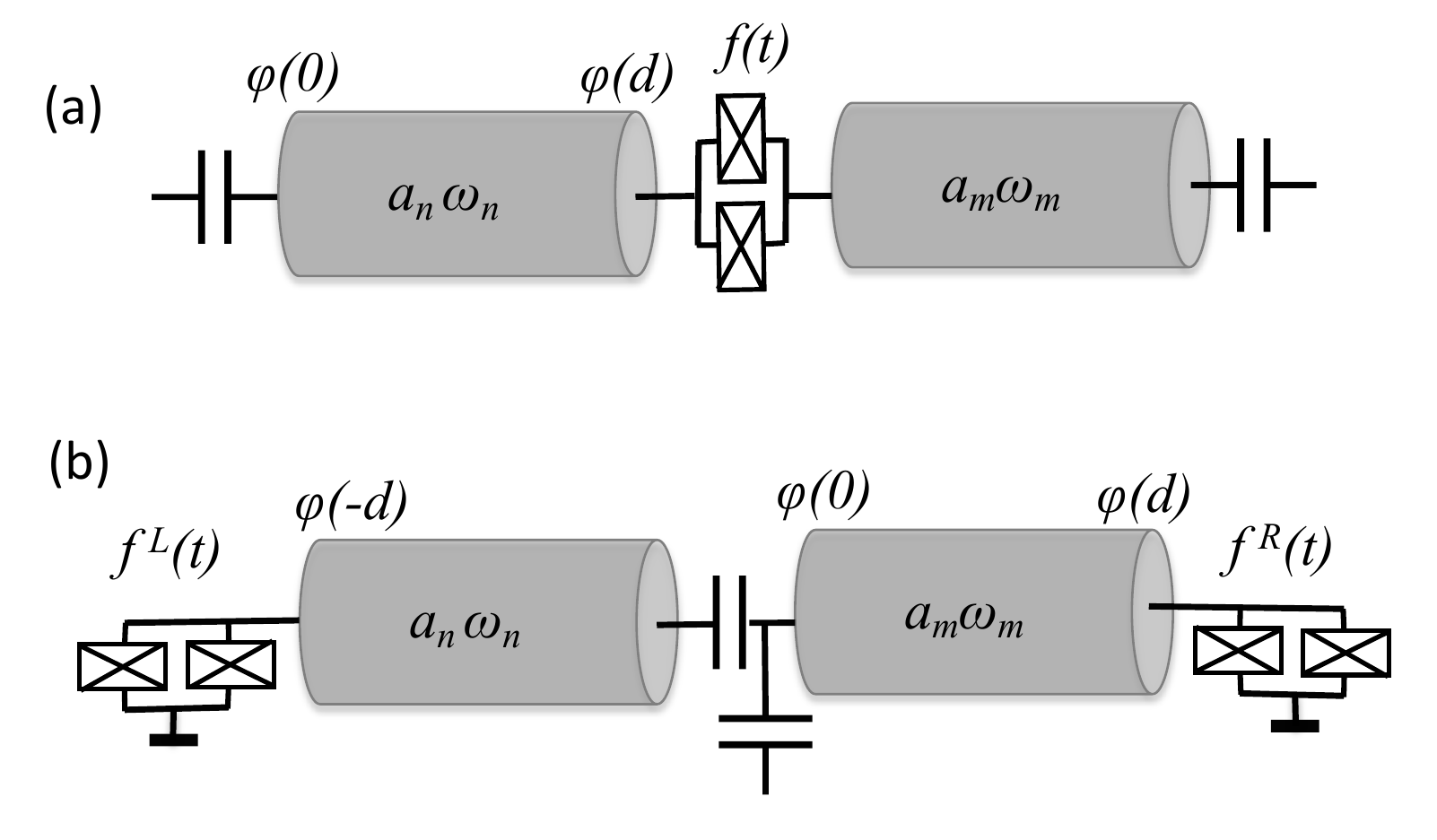}
\caption{Elementary netwoks with tunable cavities. (a) Two $\lambda/4$ resonators are coupled to same SQUID   allowing spatial separation of excited modes. (b) Two SQUIDs excite hybridized modes in two connected resonators, phase shift between pumps, $f^{R,L}(t)$, produces interference effect.}
\label{fig:cavities} 
\end{figure}

In recent experiments more complex cavity configurations where explored: $\lambda/4$ cavity connected to several SQUIDs\cite{VionPRB2014}, cavities containing SQUID arrays \cite{CasETAL2008,HakonenPNAS2013}, several cavities connected by Josephson junction network\cite{BergealNature2010,FluETAL2012,RocETAL2012}.
In Fig.~\ref{fig:cavities} we present elementary structures with two tunable cavities. Connection of two cavities to a common SQUID (panel (a)) allows one to parametrically excite spatially separated modes in different cavities; connection
of two tunable cavities as shown in panel (b) allows one to observe a parametric  interference effect produced by two pumps\cite{SvenssonLT2018,SvenssonPhD,BengtssonMS2015}. Consider performance of the latter device in more detail.

 Repeating the derivation of previous sections, one arrives at equations similar  to \Eq{eq:EOM_ndeg}, where the Kerr and pumping coefficients consist of the sum of contributions from left and right SQUIDs. Assume identical cavities, SQUIDs,  and equal modulation amplitudes and frequencies, but allow some phase shift between the pumps, $\delta f \to \delta f^{R,L} = \delta f e^{\pm i\theta/2}$. Then pumping coefficient, \Eq{eq:def_epsilon} will have the form, 
\begin{eqnarray}\label{eq:epsilon_doublesidedcavity}
\hbar\epsilon_{nm} = {\delta f\over 2} E_J\sin{F\over 2}  \left(e^{i\theta/2} s_n^R s_m^R  + e^{-i\theta/2} s_n^L s_m^L \right)
\end{eqnarray}
The $s$-coefficients are proportional to the values of the field at the cavity edges, $s_n^{R,L} \propto \phi(\pm d)$. From the symmetry of the field distribution (even or odd with respect to the origin)  it follows that  $s_n^R$ and $s_n^L$ are equal for the fundamental mode and all even modes, $n=2k$, while they have opposite signs for odd modes, $n=2k+1$. Therefore the equation in brackets has the form, 
\begin{eqnarray}\label{}
(\ldots) = s_n^R s_m^R \left(e^{i\theta/2}   + (-1)^{n+m} e^{-i\theta/2} \right).
\end{eqnarray}
and exhibits a parity effect: constructive or destructive interference effect  depending on the 
mode parity. For the degenerate resonance, $n=m$, and for the same-parity modes under the non-degenerate resonance, the pumping effect is maximum when the pumps act in phase. On the other hand, for modes with different parity the effect of the pumps is maximum when the pumps are out of phase.   

\begin{figure}[h]
\centering
\includegraphics[width=\columnwidth]{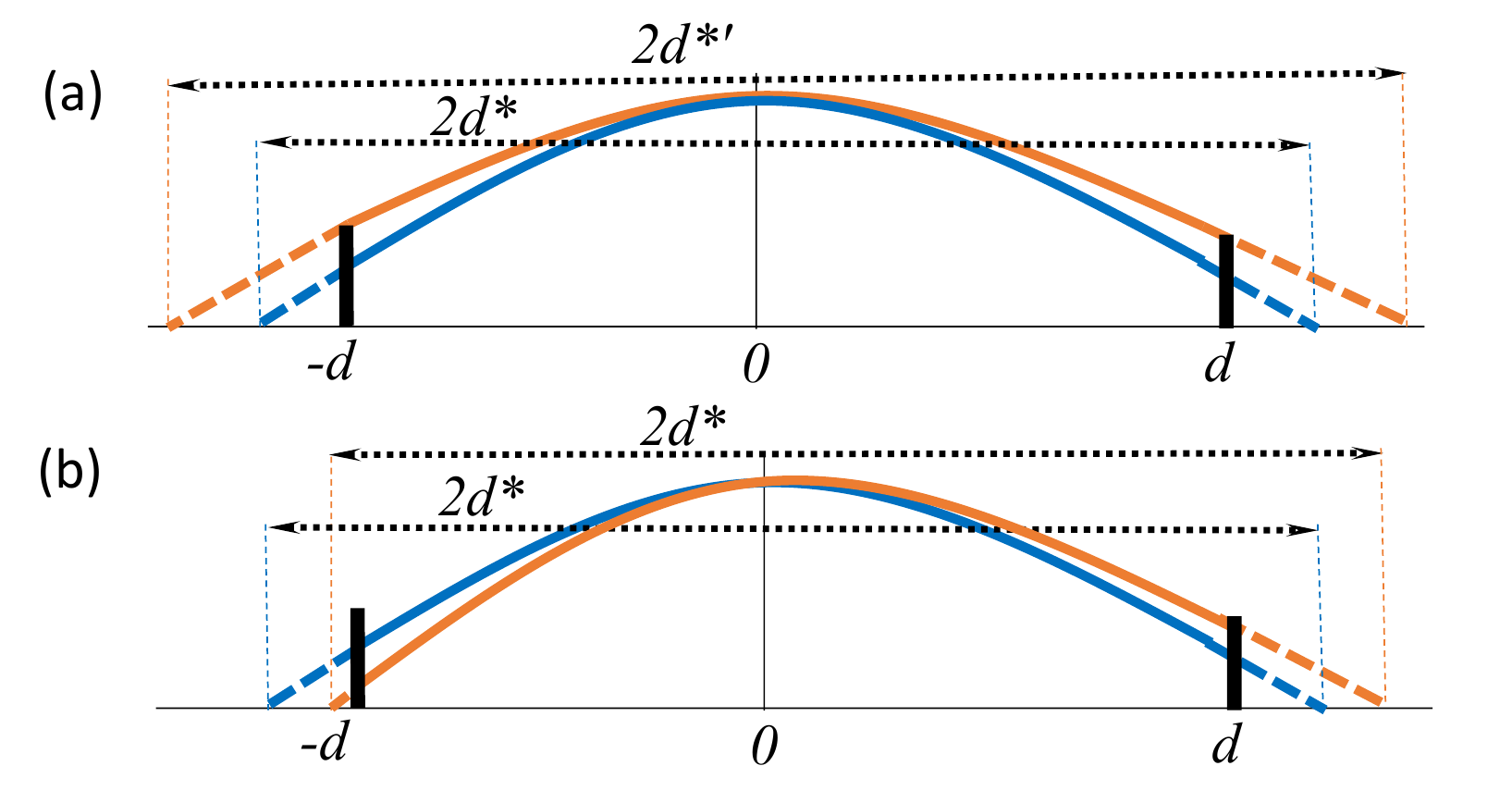}
\caption{Illustration of variation of field in the tunable cavity in terms of equivalent cavity with boundary condition 
$\phi(d^\ast) =0$. Blue continuous lines show static spatial profile of fundamental mode in two strongly coupled tunable cavities (depicted in Fig.~\ref{fig:cavities}(b)), blue dashed lines – the field in the equivalent cavity. Brown lines refer to the field under modulated boundary conditions. (a) In-phase modulation of the SQUID fluxes, $f^R(t) = f^L(t)$, is equivalent to changing length of equivalent cavity (breathing mode). (b) Out-of-phase modulation of flux, $f^R(t) = -f^L(t)$, does not change the length of equivalent cavity although changes the boundary positions (shaking mode).
}
\label{fig:shift_boundaries} 
\end{figure}

It is instructive to compare this effect with the DCE \cite{Moore1970}  in a $\lambda/2$ cavity moving in real space with boundary conditions $\phi(d)=\phi(-d)=0$, Fig.~\ref{fig:shift_boundaries}. Similar mapping on the cavity with moving mirrors  was discussed in \cite{JohanssonPRL2009,JohanssonPRA2010}. The case of in-phase modulation, $\delta f^R(t)= \delta f^L(t)$, corresponds to an antisymmetric shift of the cavity boundaries yielding change of the cavity length (breathing), as shown in panel (a). The out-of-phase  modulation, $\delta f^R(t)= -\delta f^L(t)$, is equivalent to symmetric shift of the cavity boundaries (shaking) leaving the length unchanged as shown in panel (b).  This gives natural explanation to the parametric effect in the case of escitation of individual modes and modes with equal parity. However, the fact that the parametric effect persists (for modes with different parity) when the cavity length hence eigenfrequencies do not change is surprising. This situation is analogous to the DCE with a single moving mirror \cite{WilsonNature2011}. Moreover the DCE was predicted to exist in the shaking mode of $\lambda/2$ cavity and exhibit a similar parity effect\cite{LambrechtPRL1996}.

%%%%%%%%%%%%%%%%%%%%%%%%%
\subsection{High order resonances}
The parametric resonance considered so far is the simplest and best studied resonance effect in parametrically driven systems. It is associated with temporal modulation of the system resonance frequencies. 
However, modulation of flux through the SQUID affects not just the cavity frequencies but all high order nonlinearities 
of the Josephson potential \eqref{eq:V}, $\propto\phi^N$, $N>2$. Parametric modulation of the nonlinearity coefficients gives rise to a wide class of new resonance effects that are observed in the tunable cavity and will be discussed in   
Sec.~\ref{sec:subharmonics}

%%%%%%%%%%%%%%%%%%%%%%%%%%%
%%%%%%%%%%%%%%%%%%%%%%%%%%
\section{Cavity response}\label{sec:Response}
In this section we discuss the response of parametrically pumped cavity to harmonic probe  signals. A parametric amplification is the property of the Josephson circuits\cite{YurkePRA1989} that attracted primary attention of the c-QED community, both experimentally and theoretically. The experimental work is almost exclusively done in a linear amplification regime
\cite{BergealNature2010,EichlerPRL2011,BergealPRL2012,FluETAL2012,RocETAL2012,EichlerPRL2012,YamETAL2008,VijayPRL2011,RistePRL2012,MenzelPRL2012,NakamuraAPL2013,VionPRB2014,CasETAL2008}, see also review\cite{RoyAmplification}, and a comprehensive quantum theory of linear amplification\cite{Caves1982}  was extended to the microwave domain\cite{ClerkRMP2010}. 
We will start with the linear 
amplification theory in the context of the tunable cavity. This theory applies to both the classical and quantum regimes under the degenerate and non-degenerate  resonance conditions. It is interesting to mention that most of the  observations 
made in the quantum theory of linear amplifiers\cite{Caves1982} can be found already in the classical theory, such as 
an emergence of idlers, amplification and de-amplification (squeezing) of quadratures, relations between gains, etc. Then we 
proceed to the nonlinear amplification, the regime particularly relevant in the vicinity of a parametric instability threshold, 
where already a single photon input may generate a strong classical field within the cavity. In Sec.~\ref{sec:strong_weak} we discuss  amplification of a weak signal in presence of a strong field in the cavity, which is important for the discussion of quantum noise in Sec.~\ref{sec:quantum}. Here a novel feature of  four-mode squeezing  appears under the non–degenerate resonance.  We conclude our discussion of the cavity response with studying a parametric frequency conversion.

%%%%%%%%%%%%%%%%%%%%%%%%%%%%%%
\subsection{Linear amplification}
\label{sec:linear_amplification}

\begin{figure}[h]
\centering
 \includegraphics[width=0.8\columnwidth]{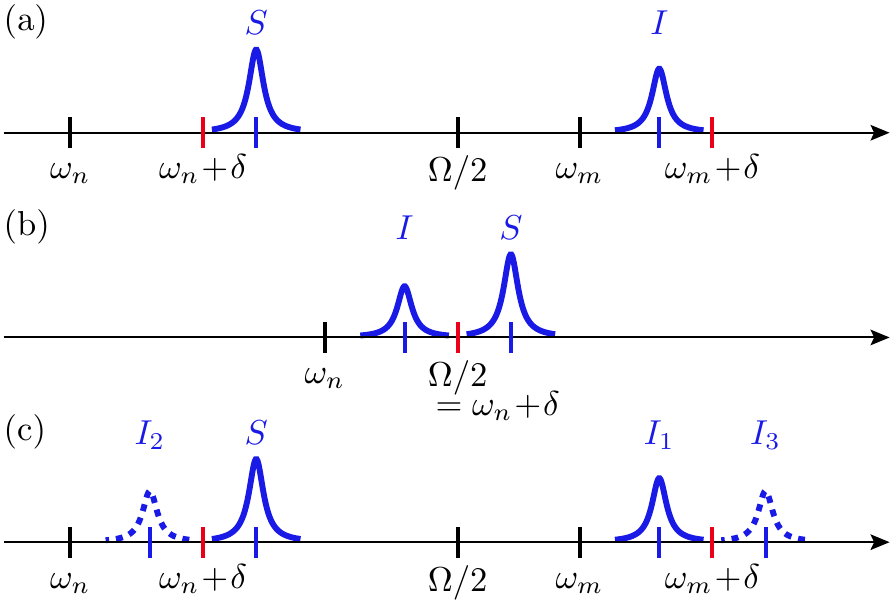}
\caption{Mode structure of amplified output field for a detuned input signal (S), 
for (a) the two-mode amplification of (a) non-degenerate parametric resonance
and (b) degenerate parametric resonance, 
and for (c) the four-mode amplification. 
Black color indicates cavity resonances, 
red color marks parametrically coupled strong field modes; 
solid blue lines indicate signal (S) and (primary) idler ($I$, $I_1$) with frequencies, $\omega_n+\delta+\Delta$ and $\omega_m+\delta -\Delta$; 
dashed blue lines indicate secondary idlers ($I_{2,3}$) with frequencies, $\omega_n+\delta-\Delta$ and $\omega_m+\delta +\Delta$.
 }\label{fig:modecoupling_diagrams}
\end{figure}

Consider the Langevin equation for nondegenerate resonance, \Eq{eq:EOM_ndeg}, and assume an incoming harmonic mode slightly detuned from rotating frame, $b_n(t) = b_n(\Delta)e^{-i\Delta t}$,  (in laboratory frame this mode has frequency $\omega = \omega_n+\delta+\Delta$). It will generate a field inside the cavity with the same frequency, $a_n(t) = a_n(\Delta)e^{-i\Delta t}$. However, because of the structure of equations in \Eq{eq:EOM_ndeg}, which connect this mode to the conjugated mode, $a_m^\dag(t)$, a field component, $a_m(t) = a_m(-\Delta)e^{i\Delta t}$, will also appear inside the cavity, with frequency $\omega = \omega_m+\delta-\Delta$ in the laboratory frame, see Fig. \ref{fig:modecoupling_diagrams}(a). This pair of modes, 
\begin{eqnarray}\label{eq:a_ndeg}
a(t) = a_n(\Delta)e^{-i\Delta t} + a_m(-\Delta)e^{i\Delta t} \,,
\end{eqnarray}
will generate a similar mode pair in the output field, which are called signal and idler. The same mode pair would be generated by an incoming mode $b_m(t) = b_m(-\Delta) e^{i\Delta t}$. 

For sufficiently weak inputs, the Kerr terms in  \Eq{eq:EOM_ndeg} can be neglected, and the equations become linear. Solving them we derive, with help of \Eq{eq:cb_general}, the linearized input-output relation,
\begin{eqnarray}\label{eq:BT}
c_n(\Delta) = u_n(\Delta) b_n(\Delta)  + v_n(\Delta) b_m^\dag(-\Delta), 
\end{eqnarray}
and similar for $c_m(\Delta)$. The coefficients in these relations have explicit form, 
\begin{eqnarray}
\label{eq:u_ndeg_linearized}
u_n(\Delta) 
&=& \frac{(\delta + \Delta - \ui \Gamma_n)(\delta - \Delta - \ui \Gamma_m) - \epsilon_{nm}^2}{(\delta + \Delta + \ui \Gamma_n)(\delta - \Delta - \ui \Gamma_m) - \epsilon_{nm}^2}  \\
\label{eq:v_ndeg_linearized}
v_n(\Delta) 
&=& \frac{2\ui \epsilon_{nm} \sqrt{\Gamma_n \Gamma_m} }{(\delta + \Delta + \ui \Gamma_n)(\delta - \Delta - \ui \Gamma_m) - \epsilon_{nm}^2}
\,.
\end{eqnarray}
According to \Eq{eq:BT}, creation and annihilation operators are mixed in the output field, which is  similar to the Bogoliubov transformation (BT) in the theory of  superfluidity and superconductivity. The Bogoliubov coefficients in \Eqs{eq:u_ndeg_linearized}-(\ref{eq:v_ndeg_linearized}) satisfy important relations,
\begin{eqnarray}\label{eq:uvproperty_ndeg}
 |u_{n}(\Delta)|^2 - |v_{n}(\Delta)|^2 &=& 1 \nonumber\\
 u_n(\Delta)v_m(-\Delta)- v_n(\Delta)u_m(-\Delta) &=& 0 \no
 v_n(\Delta) = - v_m^\ast(-\Delta) 
\,.&&
\end{eqnarray}
According to the first of these relations one can parametrize,
\begin{eqnarray}\label{eq:r}
|u_n(\Delta)| =  \cosh r_n(\Delta), \quad  |v_n(\Delta)| =  \sinh r_n(\Delta),
\end{eqnarray}
with the squeezing parameter $r_n(\Delta)$. The squeezing parameters of coupled modes are related, 
$r_m(\Delta) = r_n(-\Delta)$. Furthermore, in the quantum regime, these relations provide preservation of the bosonic commutation relations \cite{Caves1982}:  if the  Fourier harmonics of the input operators obey relations $[b_n(\Delta)\,, b_m^\dag(\Delta')] = \delta_{nm}\delta(\Delta-\Delta')$, the same is true for the output operators, $[c_n(\Delta)\,, c_m^\dag(\Delta')] = \delta_{nm}\delta(\Delta-\Delta')$. %

The amplification effect is quantified with the signal gain, 
$G_{nn}(\Delta) = \la  c_n^\dag(\Delta)c_n(\Delta)\ra / \la  b_n^\dag(\Delta)b_n(\Delta)\ra$, i.e. the ratio of the output vs input average photon numbers, and the idler gain (or cross gain),  $G_{nm}(-\Delta) = \la  c_m^\dag(-\Delta)c_m(-\Delta)\ra / \la  b_n^\dag(\Delta)b_n(\Delta)\ra$. The gains are fully characterized with the squeezing parameter,
\begin{eqnarray}\label{eq:gains_ndeg}
G_{nn}(\Delta) = \cosh^2 r_n(\Delta), \; G_{nm}(-\Delta) = \sinh^2 r_n(\Delta) \,.
\end{eqnarray}
and satisfy relations,
\begin{eqnarray}\label{eq:gain_property}
G_{nn}(\Delta) &=& 1 + G_{nm}(-\Delta)  \\
G_{mm}(\Delta) &=& G_{nn}(-\Delta), \quad G_{nm}(-\Delta) = G_{mn}(\Delta) \,, \nonumber
\end{eqnarray}

Relations \Eqs{eq:BT}-\eqref{eq:gain_property} remain formally valid for the degenerate resonance under assumption $m=n$. However, there is a difference, the signal and idlers are tightly spaced within the mode bandwidth, see Fig.~\ref{fig:modecoupling_diagrams}(b), which has physical implications for quadrature squeezing.

\begin{figure}[h]
\centering
 \includegraphics[width=0.9\columnwidth]{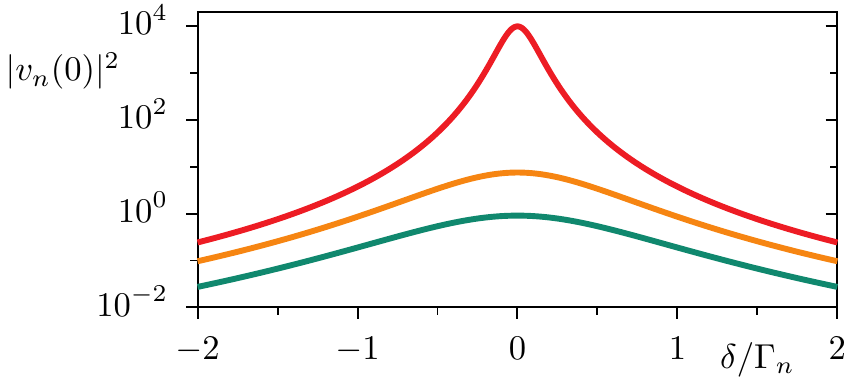}
\caption{Linear amplification of undetuned input signal, $\Delta=0$, for degenerate parametric amplifier: 
idler gain, $G^{I}(0) = |v_n(0)|^2$, vs pump detuning $\delta$,  for increasing pump strengths, 
$\epsilon_n/\Gamma_n = 0.4, 0.7, 0.99$ (from bottom to top).
}
\label{fig:lineargains_degen}
\end{figure}

To get better insight in the amplification property of the parametrically driven  cavity let us examine \Eq{eq:v_ndeg_linearized}  for the degenerate case  and on-resonance input, $\Delta = 0$, 
\begin{eqnarray}\label{v(0)}
|v_n(0)| = {2\epsilon_{n}  \Gamma_n\over \delta^2 + \Gamma_n^2 - \epsilon_n^2}\,.
\end{eqnarray}
This quantity defines the gains and  resembles the response of a damped linear oscillator driven by a force detuned by $\delta$, see Fig.~\ref{fig:lineargains_degen}. However there is an important difference:   \Eq{v(0)} refers 
to the output field rather than intrinsic field of the oscillator - the former always equals unity in case of non-parametrically driven oscillator (in absence of internal losses).  While keeping a Lorentzian shape,  the magnitude of the response (\ref{v(0)}) grows with growing pumping strength, and the width of the resonance decreases; this can be interpreted as an effective reduction of damping by parametric pumping. The full compensation of damping occurs at $\epsilon_n^2 = \Gamma_n^2 + \delta^2$, and indicates the development of parametric instability, which is known in mechanics as the parametric resonance, when the cavity intrinsic field grows without limit. Stabilization of this growth requires inclusion of the Kerr effect. 

%

%%%%%%%%%%%%%%%%%%%%%%%%%%%%%%%%%%%%%%

\subsection{Squeezing and phase sensitive amplification}
\label{sec:phase_sensitive}

The gain quantifies the absolute value of the output field. However, parametric amplification exhibits a nontrivial behavior of output quadratures, $q^c= (c+c^\dag)/\sqrt2$ and $p^c=-i(c-c^\dag)/\sqrt2$, namely amplification of certain quadratures and deamplification (squeezing) of other quadratures (for electric oscillators the quadratures correspond to voltage and current variables directly measured in experiment). To see this we
gauge out  phase factors from the $uv$-coefficients in \Eq{eq:uvproperty_ndeg},  and include them into new bosonic 
operators, giving the new BT for these operators,
\begin{eqnarray}\label{}
\tilde c_n(\Delta) = \cosh r_n(\Delta) \tilde b_n(\Delta) + \sinh r_n(\Delta) \tilde b_m^\dag(-\Delta),
\end{eqnarray}
and similar for $\tilde c_m(-\Delta)$.
The corresponding quadratures are,
\begin{eqnarray}\label{}
&& \tilde q_n^c(\Delta) = \cosh r_n(\Delta) \tilde q_n^b(\Delta) + \sinh r_n(\Delta) \tilde q_m^b(-\Delta) \no
&& \tilde p_n^c(\Delta) = \cosh r_n(\Delta) \tilde p_n^b(\Delta) - \sinh r_n(\Delta) \tilde p_m^b(-\Delta)\,. 
\end{eqnarray}
Now  we consider collective quadratures, $q_\pm(\Delta) = \left(q_n(\Delta) \pm q_m(-\Delta)\right)/\sqrt2$, and similar for $p_\pm(\Delta)$. Then for these collective quadratures we get,
\begin{eqnarray}\label{eq:collective_qp}
&& \tilde q_\pm^c(\Delta) = e^{\pm r_n(\Delta)} \tilde q_\pm^b(\Delta), \;  
\tilde p_\pm^c(\Delta) = e^{\mp r_n(\Delta)} \tilde p_\pm^b(\Delta)\,, 
\end{eqnarray}
i.e. quadratures $\tilde q_+(\Delta)$ and $\tilde p_-(\Delta)$ are amplified while  quadratures $\tilde q_-(\Delta)$ and $\tilde p_+(\Delta)$ are squeezed. 

Applying the result in \Eq{eq:collective_qp} to on-resonance input  under the degenerate resonance, where $q_-(0) = p_-(0) = 0$,  we find that  the amplification (squeezing) of $\tilde q_+ \propto q_n$ 
($\tilde p_+ \propto p_n$) refers to the signal quadratures themselves. The direction of squeezing is defined by the phases of the Bogoliubov coefficients. 

\begin{figure}[tb]
\centering
\includegraphics[width=0.9\columnwidth]{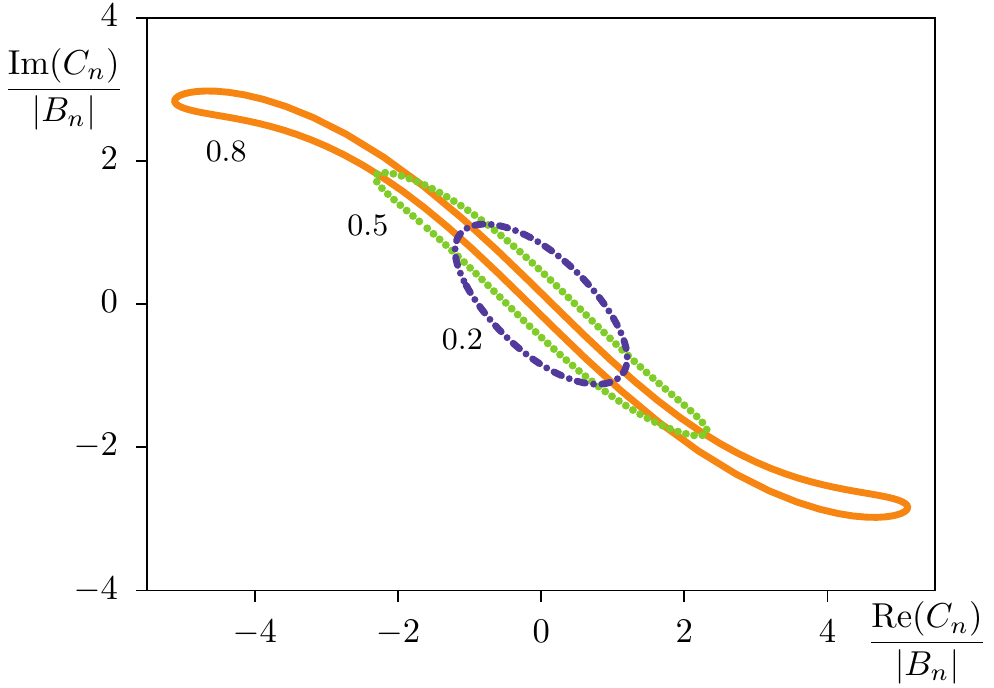}
\caption{ Anisotropy of the cavity output field in the complex $C_n$-plane under variation of input phase $\theta_B \in [0,2\pi)$, for degenerate parametric amplifier at pump strengths $\epsilon_n/\Gamma_n = 0.2, 0.5, 0.8$ (purple, green, orange), the purple curve depicts the linear regime, other curves refer to nonlinear amplification discussed in 
Sec.~\ref{sec:nonlinear_amplification}. [$\delta = 0$, $\Delta=0$, $|B_n|^2=2\Gamma_n$, $\alpha_n = \Gamma_n/100$.]
}
\label{fig:Csq_phin_deg}
\end{figure}

As it was already mentioned before, the properties of linear amplification apply to both quantum and classical fields. In the latter case we assume an input tone to be a coherent state that has non-vanishing average, $B_n=\la b_n\ra = |B_n|e^{i\theta_B} \neq 0$, where phase $\theta_B$ is referenced to the pump. This input will generate a classical intracavity field, $A_n=\la a_n\ra$, as well as a classical output field, $C_n=\la c_n\ra$, and BT can be directly formulated in terms of the classical fields. In the classical regime,  a new aspect comes to attention - the effect of the phase of the input on the squeezing direction. In Fig.~\ref{fig:Csq_phin_deg} this behaviour can be seen in the purple ellipse illustrating the dependence of the output amplitude $C_n$ on the input phase $\theta_B$. 
Another novel aspect is the input-phase dependence of the gain for on-resonance input under degenerate resonance - phase sensitive amplification.  Indeed, the BT in this case involves the input field and its complex conjugate, thus the gain includes an interference term, 
\begin{eqnarray}\label{eq:Gphase}
&& C_n(0) = \left( u_n(0) e^{i\theta_B} + v_n(0) e^{-i\theta_B}\right) |B_n(0)| \,,\no
&& G_n(0) = e^{-2r_n(0)} + 2\sinh 2r_n(0)  \cos^2(\theta_B + \eta)\,, 
\end{eqnarray}
where $\eta = (1/2){\rm arg} (u_n(0)v_n^\ast(0))$.

%%%%%%%%%%%%%%%%%%%%%%%%%%%%%%%%%
%%%%%%%%%%%%%%%%%%%%%%%%%%%%%%%%%%
\subsection{Nonlinear amplification}
\label{sec:nonlinear_amplification}

\begin{figure}[h]
\centering
 \includegraphics[width=0.9\columnwidth]{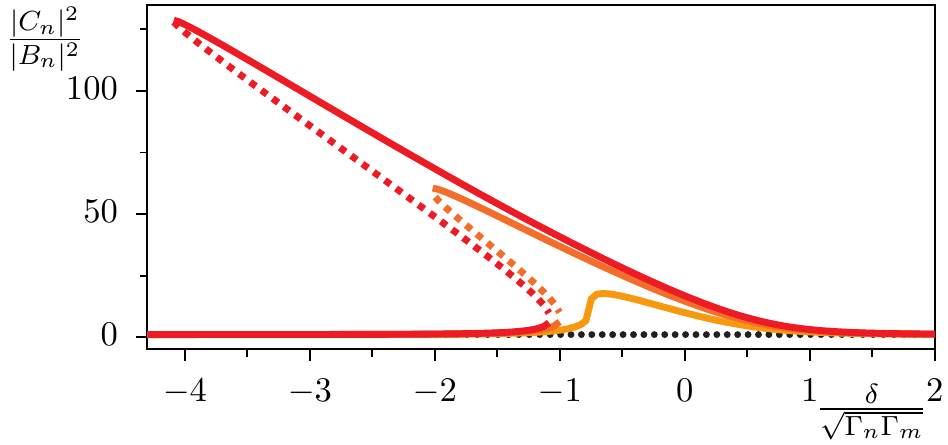}
\caption{Nonlinear gain $|C_n|^2/|B_n|^2$ vs pump detuning $\delta$,
for the non-degenerate parametric amplifier and on-resonance input, $\Delta=0$,
with $\epsilon_n/\Gamma_n = 0.8, 0.9, 0.95$ (from bottom to top); 
black dotted line refers to the Duffing limit $\epsilon_n = 0$.
[$|B_n|^2 = 2\Gamma_n$, $\theta_B=0$, $B_m = 0$, $\alpha_n/\Gamma_n = 0.01$, $\alpha_n=\alpha_m$, $\Gamma_n=\Gamma_m$.] }
\label{fig:nlin_gain_ndeg}
\end{figure}

With increasing input power the intracavity field becomes so strong that the Kerr effect can no longer be neglected.  The same is true even for weak inputs at large pumping intensity close to the instability threshold. In this regime the Kerr effect leads  to an appreciable shift of the resonance frequency, $\alpha_j|A_j|^2 \gtrsim \sqrt{\epsilon_{nm}^2-\Gamma_n\Gamma_m}$.  

For classical inputs under non-degenerate resonance the linear response theory can be straightforwardly generalized to  the nonlinear case.  To this end we replace in the Langevin equations, \Eq{eq:EOM_ndeg}, the field operators $a_j$ with classical amplitudes $A_j$, and $a_j^\dag a_j$ with $|A_j|^2$ in Kerr terms. Repeating the derivation we arrive at the same BT as in \Eqs{eq:BT} - (\ref{eq:v_ndeg_linearized}), but with the pump detuning $\delta$ being replaced  with
\begin{eqnarray}\label{zeta_ndeg}
\delta \to \zeta_{n} = \delta + \alpha_{n} |A_{n}|^2 + 2\alpha_{nm} |A_{m}|^2.
\end{eqnarray}
These generalized BT equations, however, do not provide explicit solution to the problem since they contain intracavity amplitudes, $A_n$, that are to be found self-consistently, 
\begin{eqnarray}\label{eq:nonlinearA}
(\Delta + \zeta_n + i\Gamma_n) A_n(\Delta) + \epsilon_{nm} A_m^\ast(-\Delta) = \sqrt{2\Gamma_n} B_n(\Delta) . \no
\end{eqnarray}
In spite of this complication the Bogoliubov coefficients turn out to still obey  \Eq{eq:uvproperty_ndeg}, hence the 
{\em nonlinear} gains obey the same  relations as given by \Eq{eq:gain_property}\cite{Wustmann2017}. 

The nonlinear gain as function of pump detuning is illustrated in Fig. \ref{fig:nlin_gain_ndeg} for on-resonance input, $\Delta=0$. This gain resembles the response of a nonlinear Duffing oscillator, where the maximum value is shifted from $\delta=0$ due to the Kerr effect and grows with increasing pump strength. Similar to the Duffing oscillator, a bistability region exists at red detuning, which appears at increasingly small input level and occupies larger $\delta$-interval when the pump strength increases. In addition, the resonance width becomes vanishingly small, due to effective reduction of damping. This behavior is a precursor of the transition to the regime of self-sustained parametric oscillation discussed in the next Sec.~\ref{sec:oscillation}.  One can compute the maximum gain value achieved at the instability threshold, 
$\epsilon_{nm} = \sqrt{\Gamma_n\Gamma_m}$, at $\delta =0$\cite{Wustmann2017},
\begin{eqnarray}\label{Gnonlinear}
G_{nn}(0) \approx G_{mn}(0) = {2|A_n|^2\Gamma_n\over |B_n|^2 } \gg 1 \,,
\end{eqnarray}
where the intracavity field is given by equation,
\begin{eqnarray}\label{eq:A1approx@threshold}
|A_n|^6 = {2\over (1-\Gamma_n/\Gamma_m)^2} \left({\Gamma_n\over\alpha_n}\right)^2 
{|B_n|^2\over \Gamma_n} \,.
\end{eqnarray}
It follows from these equations that the maximum nonlinear gain has a non-analytic dependence on the input power, 
$G_{nn}(0) \sim 1/|B_n|^{2/3}$, and diverges at small power. Furthermore, its value is controlled by the parameter 
$\Gamma_n/\alpha_n$, which also means that this parameter constraints the squeezing parameter $r_n(0)$ by virtue of 
\Eq {eq:r}, and therefore defines the maximum level of squeezing.

\begin{figure}[tb]
\centering
\includegraphics[width=\columnwidth]{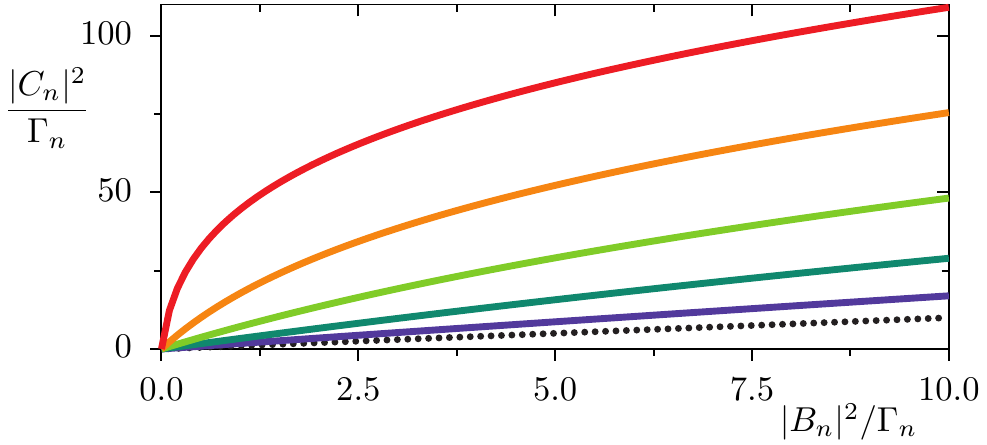}
\caption{Output power $|C_n|^2$ vs input power $|B_n|^2$,
for the degenerate parametric amplifier with on-resonance input signal, $\Delta=0$, 
and $\epsilon_n/\Gamma_n = 0, 0.2, 0.4, 0.6, 0.8, 1.0$ 
(from bottom to top); the dotted line refers to the Duffing limit $\epsilon_n = 0$.
($\delta/\Gamma_n = 0.5$, $\theta_B=\pi/2$, $\alpha_n/\Gamma_n = 0.01$.)}
\label{fig:Csq_vs_Bsq_deg}
\end{figure}

In the case of degenerate parametric resonance, a generalization of \Eqs{eq:BT} - \eqref{eq:v_ndeg_linearized} to the nonlinear regime does not apply for detuned inputs.
 The reason is that the signal and idler frequencies lie close to each other within the mode bandwidth, see Fig.~\ref{fig:modecoupling_diagrams},  and their interference makes the Kerr term time dependent. This implies that the nonlinear amplification of detuned signals is nonstationary. This difficulty does not exist for on-resonance input, where the stationary nonlinear BT equations are valid, and the response is similar to the one depicted in Fig.~\ref{fig:nlin_gain_ndeg}. 
The dependence of the output power on the input for different pumping strengths is illustrated in Fig. \ref{fig:Csq_vs_Bsq_deg}. The differential gain at small input power grows without limit when the pumping strength approaches the instability threshold, and it saturates at high inputs. The phase dependence of the nonlinear gain also deviates from the linear behavior \cite{WustmannPRB2013}, as is shown in Fig.\ref{fig:Csq_phin_deg}. It is because of the phase dependence of the Kerr frequency shifts,  \Eq{zeta_ndeg}, reflecting anisotropy of the intracavity field amplitude. 

%

%%%%%%%%%%%%%%%%%%%%%%%%%%%%%%%%%%%%%%%%%%%%%%%%%%%
\subsection{Linear response in presence of strong field}
\label{sec:strong_weak}
Here we consider amplification of a weak detuned signal in the presence of strong on-resonance field in the cavity. The latter may be generated by an on-resonance input or, if the cavity is pumped above the parametric instability threshold, by parametric oscillation, as discussed later in Sec.~\ref{sec:oscillation}. The results will be used in Sec.~\ref{sec:homodyne} to evaluate the quantum noise and signal-to-noise ratio.

%%%%%%%%%%%%%%%%%%%%%%%%%%%%%%%%%
\subsubsection{Degenerate resonance: two-mode amplification}
\label{sec:strong_weak_deg}%

Suppose that a weak detuned signal is applied on top of a strong on-resonance signal, 
$B(t) = B + b(\Delta) e^{-i\Delta t}$ (here we suppress the mode index for brevity, and use small letter for weak field since it can refer to classical field as well as quantum mode). The term "weak" refers to the filed whose contribution to the Kerr effect can be neglected. This will generate an intracavity field, $A(t) = A + a(\Delta) e^{-i\Delta t}$, with $A$ given by \Eq{eq:nonlinearA} with $m=n$ and $\Delta=0$, and with a small correction $a$,
$\la a^\dag a\ra \ll |A|^2$ that satisfies a linearized Langevin equation,
\begin{eqnarray}\label{eq:EOM_A_linearized}
(\znl + \Delta + \ui\Gamma) a(\Delta)  + \tilde\epsilon a^\dag(-\Delta)
=  \sqrt{2 \Gamma} b (\Delta)
\,,
\end{eqnarray}
where $\znl = \delta + 2\alpha |A|^2$,  and $\enl = \epsilon + \alpha A^2$. This equation is similar to linearized \Eq{eq:EOM_ndeg}, it has the same two-mode structure combining signal and idler, but the detuning here is Kerr-shifted due to strong field, and the pump intensity is also renormalized. These modifications do not affect the structure of the BT, \Eq{eq:BT}, and the form of the Bogoliubov coefficients in \Eqs{eq:u_ndeg_linearized}-(\ref{eq:v_ndeg_linearized}), as long as corresponding modifications,   $\delta \to \znl$, $\epsilon \to \enl$,  are included (note that since the effective pump strength is complex the change, $\epsilon^2 \to |\tilde\epsilon|^2$, is to be made). Correspondingly, the general properties of parametric amplification, 
\Eqs{eq:uvproperty_ndeg}-(\ref{eq:gain_property}), are preserved although the gains become different.

The way the effective pumping strength, $\tilde\epsilon$, is modified indicates that the strong intracavity field acts as an additional parametric pump - current pump. This mechanism involving a four wave mixing works even in absence of flux pumping, $\epsilon=0$, as soon as a strong on-resonance signal is injected in the cavity. In fact, it is this mechanism of current pumping that was used in many realizations of quantum limited parametric amplifiers\cite{BergealNature2010,EichlerPRL2011,BergealPRL2012,FluETAL2012,RocETAL2012,EichlerPRL2012}.

%%%%%%%%%%%%%%%%%%%%%%%
\subsubsection{Non-degenerate resonance: four-mode amplification}
\label{sec:strong_weak_ndeg}

Proceeding to the non-degenerate resonance we encounter a completely different situation. A strong applied  input to 
either of the two parametrically coupled modes will generate  a strong intracavity field in both modes. As it was found in the 
previous section, these fields will act as additional parametric pumps that will generate additional idlers for a weak detuned 
signal \cite{Wustmann2017}.  As a result, amplification of a weak signal is generally accompanied by three idlers as 
depicted in Fig.~\ref{fig:modecoupling_diagrams}(c). 

While input detuning could be included suppose for simplicity an on-resonance  strong field being applied to either (or 
both) of modes $n,m$, and additionally a weak detuned signal is  applied.  Then a total stationary  intracavity field in each 
mode will contain one strong and two weak components, 

$A_j(t) =  A_{j} + \sum_{\pm} a_j(\pm\Delta) e^{\mp i \Delta t}$.  The strong components are described with \Eq{eq:nonlinearA} (setting $\Delta=0$), and the weak components satisfy the linearized equations,
\begin{eqnarray}\label{eq:EOM_ndeg_linear}
&& ( \overline\zeta_{n} \pm \Delta  + i \Gamma_n) a_n(\pm\Delta) + 2\alpha_{nm} A_{n}A_{m}^\dag a_m(\pm \Delta) \nonumber\\
& + & \alpha_n A_{n}^2 a_n^\dag(\mp\Delta) + \overline\epsilon_{nm} a_m^\dag(\mp\Delta)  
   = \sqrt{2\Gamma_{n}} \, b_n(\pm\Delta),  
\end{eqnarray}
and similar for the $m$-th mode. Here the shifted detuning and renormalized pump strength read,
\begin{eqnarray}
\label{eq:zetas_abovethresh}
&&  \overline\zeta_{n} = \delta +  2 \alpha_n |A_{n}|^2 + 2 \alpha_{nm} |A_{m}|^2  \no
%%%
\label{eq:epsilon_abovethresh}
&& \overline\epsilon_{nm} = \epsilon_{nm} + 2\alpha_{nm} A_{n} A_{m}
\,. 
\end{eqnarray}
As we see from \Eq{eq:EOM_ndeg_linear} the strong intracavity fields not only contribute to the pump-induced intermode coupling, $a_n(\Delta) \leftrightarrow a_m(-\Delta)$, but generate {\em intramode} coupling,  $a_n(\Delta) \leftrightarrow a_n(-\Delta)$, and also generate {\em intermode} frequency conversion, $a_n(\Delta) \leftrightarrow a_m(\Delta)$. One can also see that the two latter mechanisms depend entirely on the strong intracavity fields.

Given the four-mode structure of \Eq{eq:EOM_ndeg_linear}, the BT also acquires the four-mode structure,
\begin{eqnarray}\label{eq:BT_fourmode}
c_i(\Delta) = U_{ij}(\Delta) b_j(\Delta) + V_{ij}(\Delta) b_j^\dag(-\Delta) 
\end{eqnarray}
($i, j = n, m$). Here the scalar Bogoliubov $uv$-coefficients of the two-mode amplification are replaced with matrices, whose elements determine the gains of signal and idlers. Similar to linear amplification, \Eq{eq:BT_fourmode} is also valid in the quantum regime, however, 
in order to guarantee bosonic properties of the output operators, the multimode Bogoliubov matrices have to respect constraints imposed on their unitary equivalent diagonal forms\cite{BraunsteinPRA2005,TrepsD2010}. 
Explicit analytical calculation of Bogoliubov $UV$-matrices and their diagonalization is a cumbersome task. However, this can relatively easily be done within a model of balanced modes\cite{FabreJdP1989}.

%%%%%%%%%%%%%%%%%%%%%%%%%%
\subsubsection{Balanced mode model}
\label{Sec_analysis_4mode}

In this model one assumes equal Kerr coefficients and  damping rates for both modes, $\alpha_n=\alpha_m$, 
$\Gamma_n=\Gamma_m$. One has to admit that this model is rather artificial since real cavity parameters are strongly 
frequency dependent. As shown in\cite{Wustmann2017} diagonalization of the BT \eqref{eq:BT_fourmode} within the balanced mode model  is achieved with unitary transformation, 
\begin{eqnarray}\label{fraku}
{\mathfrak U} =  {1\over \sqrt 2} \left(
\begin{array}{cc}
 e^{i \psi/2} &  e^{i \psi/2}  \\
e^{-i \psi/2} & -e^{-i \psi/2}  
\end{array}
\right) \,,
\end{eqnarray}
where the phase $\psi = \theta_n - \theta_m$ is related to the phases of the strong intracavity fields, $A_{j} = |A_{j}|e^{i\theta_j}$. 
The unitary transformation in \Eq{fraku} defines the "supermode" operators, $b_\sigma(\Delta) = {\mathfrak U}_{\sigma j}^\dag b_j (\Delta)$, $\sigma= \pm$.
In the supermode basis, the BT in \Eq{eq:BT_fourmode} splits into two independent equations, whose structure reproduces the BT for the degenerate parametric resonance,
\begin{eqnarray}\label{BTsigma}
 c_\sigma (\Delta) = u_\sigma(\Delta)  b_\sigma (\Delta) + v_\sigma(\Delta)  b_\sigma^\ast(-\Delta) \,, 
\end{eqnarray}
with coefficients,
\begin{eqnarray}
\label{eq:uBT_pmbasis}
u_{\sigma}(\Delta) &=& { (\zeta_\sigma +\Delta -  i\Gamma)(  \zeta_\sigma -\Delta   - i\Gamma)    - |\epsilon_\sigma |^2  \over (\zeta_\sigma +\Delta +  i\Gamma)(  \zeta_\sigma -\Delta   - i\Gamma)    - |\epsilon_\sigma |^2 } \nonumber\\
v_{\sigma}(\Delta) &=& {  2i\Gamma\epsilon_\sigma  \over (\zeta_\sigma +\Delta +  i\Gamma)(  \zeta_\sigma -\Delta   - i\Gamma)    - |\epsilon_\sigma |^2} 
\,,
\end{eqnarray}
where $\zeta_\sigma  = \delta + (4 + 2\sigma)\alpha_n  |A_n|^2$,  and
$\epsilon_\sigma = \sigma \epsilon_{nm} + (2\sigma +1)\alpha_n A_{n}A_{m}$
(equation $|A_n| = |A_m|$ holds in the balanced mode model). 
Since the structure of these equations exactly reproduces the one of \Eqs{eq:u_ndeg_linearized}-(\ref{eq:v_ndeg_linearized}), the properties \Eqs{eq:uvproperty_ndeg} hold for the supermodes, and therefore the supermode output operators are bosonic. This is also true for the original output operators due to the unitarity of the transformation matrix,  \Eq{fraku}. Using this method, the gains for amplified weak signals and idlers are analyzed in Ref.~\cite{Wustmann2017}.

%%%%%%%%%%%%%%%%%%%%%%%%%%%%%%%%%%%%%%%%
%%%%%%%%%%%%%%%%%%%%%%%%%%%%%%%%%%%%%
\subsection{Frequency conversion}
\label{Conversion}
In this section we explore a different parametric regime producing frequency conversion. Suppose the SQUID is modulated with frequency equal to a {\it difference} between two cavity eigenmodes,
$\Omega = \omega_n - \omega_m + 2\delta$, ($\omega_n>\omega_m$).  The Langevin equation is derived in this case in the reference frame rotating with frequencies $\omega_n+\delta$ and $\omega_m-\delta$, and following the derivation in Sec.~\ref{sec:Resonance}, we get\cite{Wustmann2017},
\begin{eqnarray}
\label{eq:EOM_upconv}
 i \dot a_n &+& 
 (\delta  + \ui \Gamma_n  + \alpha_n \,a_n^\dag a_n + 2 \alpha_{nm} \,a_m^\dag a_m ) a_n  + \epsilon_{nm} a_m
  \nonumber \\  
 & = & \sqrt{ 2\Gamma_{n}} \,b_n(t)\,,
\end{eqnarray}
and a similar equation for $a_m$ where change $\delta\to -\delta$ is made. The major qualitative difference of this equation from the Langevin equation  {\eqref{eq:EOM_ndeg} is the presence of the amplitude rather than the conjugated amplitude of the second mode. As a result, there is no parametric amplification, but a beam splitter effect, where an input mode with frequency close to $\omega_n$ is converted to an output mode with frequency close to $\omega_m$.

Theory of a nonlinear frequency conversion is technically similar to the theory of nonlinear amplification in Section \ref{sec:nonlinear_amplification}. Consider a classical input in the form of equally detuned harmonic signals in each mode, 
$B_{n,m}(t) = B_{n,m}(\Delta) e^{-i\Delta t}$. Then the input-output relation takes the form of two-mode scattering equations,
\begin{eqnarray}\label{eq:CBrelation_upconv}
C_n(\Delta) &=& S_{nn}(\Delta) B_n(\Delta) +   S_{nm}(\Delta) B_m(\Delta)\,,
\end{eqnarray}
where the coefficients from a unitary scattering matrix, $\hat S^\dag = \hat S^{-1}$. Their explicit form is similar to the Bogoliubov $uv$-coefficients,
\begin{eqnarray}\label{Scoefficients}
S_{nn}(\Delta) &=&   {(\zeta_n + \Delta - i\Gamma_{n}) (\zeta_m + \Delta+i\Gamma_m) - \epsilon_{nm}^2 \over 
(\zeta_n+ \Delta+i\Gamma_n)(\zeta_m + \Delta+ i\Gamma_m) - \epsilon_{nm}^2 }  \no
S_{nm}(\Delta) &=& {2i \epsilon_{nm}\sqrt{\Gamma_n\Gamma_m}\over (\zeta_n+ \Delta+i\Gamma_n)(\zeta_m + \Delta+ i\Gamma_m) - \epsilon_{nm}^2 }\,.
\end{eqnarray}
Here $\zeta_j$ are defined slightly differently compared to the amplification case,
\begin{eqnarray}\label{eq:zetas_upconv}
\zeta_n =  \delta + \alpha_n |A_n|^2 + 2\alpha_{nm} |A_m|^2 \nonumber\\
\zeta_m = -  \delta + \alpha_m |A_m|^2 + 2\alpha_{nm} |A_n|^2\,.
\end{eqnarray}
but $\epsilon_{nm}$ is given by the same equation (\ref{eq:def_epsilon}). We note that \Eqs{Scoefficients} do not give an explicit solution to the problem since they contain intracavity field amplitudes that are to be computed self-consistently. 
Since there is no amplification effect in this case, a sufficiently small input signal produces a small intracavity field, and the scattering matrix can be linearized. In this linearized form the results in \Eqs{eq:CBrelation_upconv}-(\ref{Scoefficients})
can be extended to the quantum regime. 

\begin{figure}[h]
\includegraphics[width=\columnwidth]{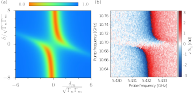}%
\caption{ Linear frequency conversion as function of input signal detuning $\delta_m = \Delta-\delta$ 
and pump detuning $\delta$. (a) Reflection coefficient $|S_{mm}(\delta_m)|^2$ quantifies response in the input mode, it exhibits an avoided crossing of a loss resonance. 
(b) Measured reflection phase, ${\rm arg} \, (S_{mm})$, as function of signal and pump frequencies \cite{SvenssonPRB2017}.
[In (a): $\Gamma_{n0}=3\Gamma_{m0}$, 
$\Gamma_{m} = 1.8\Gamma_{m0}$,
$\Gamma_{n} = 4\Gamma_{m0} = 4/3 \Gamma_{n0}$,
and $\epsilon_{nm} = 2\sqrt{\Gamma_n \Gamma_m}$].
}
\label{fig:lineargain_dw1_dpD_internalloss__frconv}
\end{figure}

In Fig.~\ref{fig:lineargain_dw1_dpD_internalloss__frconv}(a)
we show the linear reflection spectrum,  $|S_{mm}|^2$, of the parametric conversion process versus input detuning from mode frequency $\delta_m$: $\omega-\omega_m =-\delta+\Delta$, and pump detuning. Internal losses are included in the numerics.
If the pump is detuned far away from the resonance, $|\delta| \gg \sqrt{\Gamma_n\Gamma_m}$, the spectrum is dominated by a loss resonance centered at $\delta_m = 0$. Close to the parametric resonance, 
$\delta < \sqrt{\Gamma_n\Gamma_m}$, the intermode coupling appears as an avoided crossing  that is accompanied 
by the emergence of the converted signal (i.e. finite $S_{mn}$, not shown). 

The quantum frequency conversion was observed in several c-QED experiments\cite{AumentadoNaPh2011,AbdoPRL2013,AumentadoAPL2015,BengtssonMS2015} including both device configurations in Fig.~\ref{fig:cavities}. In Ref.\cite{SvenssonPRB2017} this effect was used to characterize a high order mode unaccessible for direct measurement, see Fig.~\ref{fig:lineargain_dw1_dpD_internalloss__frconv}(b). 

\begin{figure}[h]
\includegraphics[width=0.9\columnwidth]{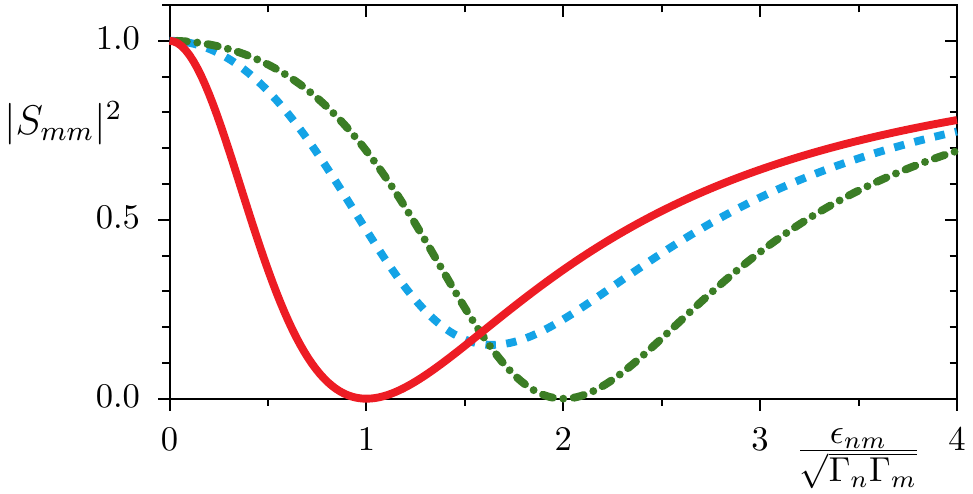}%
\caption{ Reflection coefficient $|S_{mm}(\Delta)|^2$ vs pump strength $\epsilon_{nm}$, for different pump and signal detunings: solid (red) line: $\delta = \Delta = 0$ with full conversion at $\epsilon_{nm}/\sqrt{\Gamma_n \Gamma_m} = 1$; dash-dotted (green) line: $\delta = \sqrt{3} (\Gamma_n - \Gamma_m)/2$, $\Delta = \sqrt{3} (\Gamma_n + \Gamma_m)/2$ with full conversion at $\epsilon_{nm}/\sqrt{\Gamma_n \Gamma_m} = 2$; 
dashed (blue) line: $\delta = 0$, $\Delta = (\Gamma_n + \Gamma_m)/2$.
[$\Gamma_{n0}=3\Gamma_{m0}$,  $\Gamma_{m} = \Gamma_{m0}$, $\Gamma_{n} = \Gamma_{n0}$.]
}
\label{fig:Conversion_efficiency}
\end{figure}

A full reciprocal conversion between the modes is possible in the absence of internal losses. The criterion is given by the zero reflection coefficient,  $|S_{mm}|^2=0$. The corresponding conditions for linear conversion read, 
\begin{eqnarray}\label{full_conversion}
\epsilon_{nm}^2 = \Gamma_n\Gamma_m \left( 1+ {4\delta^2\over (\Gamma_n - \Gamma_m)^2}\right), \;\;
\Delta =  {\Gamma_n + \Gamma_m \over \Gamma_n - \Gamma_m}\,\,\delta\,. \nonumber
\end{eqnarray}
It is instructive to compare these equations to the ones for the parametric instability in the amplification regime, see below \Eqs{eq:Delta_thresh} and (\ref{eq:delta_thresh}): both criteria coincide at the zero pump detuning, $\delta=0$.  At finite pump detuning full conversion is still possible, but in this case the input must be detuned accordingly. The efficiency of the frequency conversion at different pump strengths is illustrated in Fig.~\ref{fig:Conversion_efficiency}. 

\begin{figure}[h]
\includegraphics[width=\columnwidth]{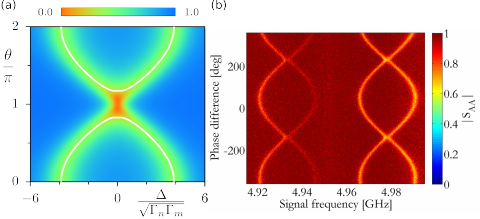}
\caption{Interference effect in frequency conversion in  coupled cavities shown in Fig.~\ref{fig:cavities}(b). 
a) Calculated reflection coefficient in \Eq{Scoefficients}, at $\delta=0$ and pumping strength replaced with  
$\epsilon_{nm}(\theta) = \epsilon_{nm}(0) \sqrt{1+\cos\theta}/\sqrt{2}$ according to \Eq{eq:epsilon_doublesidedcavity}, 
as function of signal detuning and pump phase difference (internal losses are included); white line marks the positions of the split resonance.  
(b) Measurement data adopted from\cite{BengtssonMS2015}, courtesy of A. Bengtsson (note the doubled range $\theta$, and the wide range of detuning, such that both $|S_{mm}|$ and $|S_{nn}|$ are included).
[In (a): $\Gamma_{n0}=3\Gamma_{m0}$, 
$\Gamma_{m} = 1.8\Gamma_{m0}$,
$\Gamma_{n} = 4\Gamma_{m0} = 4/3 \Gamma_{n0}$,
$\delta = 0$, 
and $\epsilon_{nm}(0) = 4\sqrt{\Gamma_n \Gamma_m}$.]
}
\label{fig:conversion_modulation}
\end{figure}

The interference effect discussed in Sec.~\ref{sec:2cavity}, \Eq{eq:epsilon_doublesidedcavity}, was tested experimentally \cite{BengtssonMS2015} by measuring frequency conversion in the device depicted in Fig.~\ref{fig:cavities}(b). The theoretical result for the reflection coefficient computed with \Eq{Scoefficients}  is shown in Fig.~\ref{fig:conversion_modulation}(a). Here the phase-dependent pumping strength, $\epsilon_{nm}(\theta)$ is taken in the form of \Eq{eq:epsilon_doublesidedcavity},  $\delta=0$, and internal losses are included; the white lines indicate the positions of the split resonance given by the extrema of \Eq{Scoefficients}. The result of the measurement of the reflection coefficient at $\delta=0$ is presented in Fig.~\ref{fig:conversion_modulation}(b); here we see the pump-phase dependent split resonances of not one but of both parametrically coupled modes of the device. These modes result from the hybridization of nearly degenerate modes of individual cavities, where the small distance between the hybridized modes is due to the relatively weak coupling of the cavities.

%%%%%%%%%%%%%%%%%%%%%%%%%%%%%%%%%%
%%%%%%%%%%%%%%%%%%%%%%%%%%%%%%%%%%
\section{Parametric oscillations}
\label{sec:oscillation}

As it was mentioned in Sec.~\ref{sec:linear_amplification} the linear response exhibits a divergence at pumping strength, 
$\epsilon_{nm} = \sqrt{\Gamma_n\Gamma_m}$ at zero pump detuning, $\delta=0$. This divergence exists in the presence of damping and therefore cannot be suppressed by increasing losses. The mechanism of removing the divergence is related to the Kerr effect - a shift of the cavity frequency by the energy of the field stored in the cavity\cite{Landau}. The state of the self-sustained oscillations that develops above the parametric threshold is therefore essentially nonlinear. It is characterized, away from the threshold, by a large average amplitude of the intracavity field, $A = \la a\ra$, which can be treated classically.
Transition from the ground state, $\la a\ra =0$, below the threshold to the state, $\la a\ra \neq 0$ above the threshold can be understood as a second order phase transition\cite{Haken} with  $A$ playing the role of  complex order parameter.
The divergence of the linear response at the threshold can then be understood as the result of critical fluctuations. 

In this section we present a classical theory of parametric oscillations under both degenerate and non-degenerate resonance. Full quantum analysis of the parametric oscillations is the subject of ongoing research. We present some results on small quantum fluctuations of the degenerate oscillator in the following  Sec.~\ref{sec:quantum}, an interpretation of the degenerate oscillations as quantum cat states will be discussed in Sec.~\ref{sec:cats}. 

\subsection{Degenerate oscillations}
\label{sec:oscillation_deg} 

%%%%%%%%%%%%%%%%%%%%%%%%%%%%%%%%
\subsubsection{Quasiclassical description}
\begin{figure}[h]
\centering
\includegraphics[width=0.7\columnwidth]{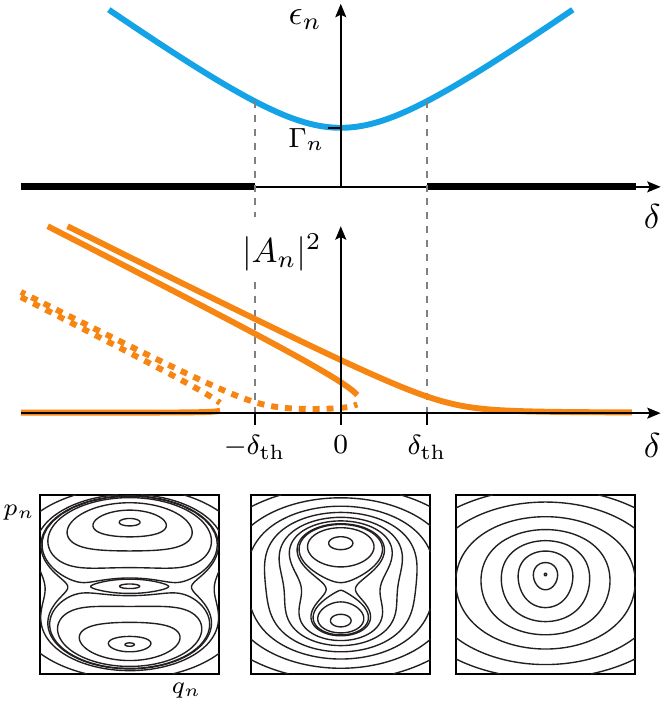}
\caption{
Degenerate parametric oscillation. Upper panel: blue line indicates  
threshold of instability, $\epsilon_n(\delta)$; bold black line indicate stability region of the ground state. Middle panel: 
stationary response to applied on-resonance signal vs detuning, intensity of the response indicates the one of the self-
sustained oscillations split by the input; solid lines indicate stable solutions born at the edge $\delta=\delta_{th}$, dashed 
lines indicate unstable solutions born at the edge $\delta=-\delta_{th}$. Lower panel: phase portraits of oscillations at 
different detuning:  stable ground state at blue detuning, $\delta>\delta_{th}$, bistable excited states at 
$-\delta_{th}<\delta<\delta_{th}$, and excited states coexisting with the stable ground state at red detuning, 
$\delta<-\delta_{th}$.
}
\label{Fig:Degenerate_th}
\end{figure}

In this section we follow analysis of Ref.\cite{WustmannPRB2013}. Stability analysis of \Eq{eq:EOM_deg} confirms that the ground state is unstable within the interval of pump detuning, 
\begin{eqnarray}\label{eq:threshold}
|\delta| < \delta_{th}(\epsilon_n) = \sqrt{\epsilon_n^2 - \Gamma_n^2},
\end{eqnarray}
and stable excited stationary states exist at $\delta < \delta_{th}$.  These oscillatory states have frequency 
$\omega=\Omega/2$, and amplitude,    
\begin{eqnarray}\label{eq:A_osc_deg}
  |A_n|^2 =  {-\delta + \sqrt{\epsilon_n^2 - \Gamma_n^2} \over \alpha_n} \,,
 \end{eqnarray}
and have two-fold degeneracy with respect to the phases, 
\begin{eqnarray}\label{eq:theta_osc_deg}
{\rm arg}\, A_n = \theta_{n}, \theta_{n} + \pi; \;   2\theta_{n} = \arcsin{\Gamma_n\over \epsilon_n} \in \left({\pi\over 2} , \pi\right) . \;\;
%\sin(2\theta_n) = {\Gamma_n\over \epsilon_n}, \;\;
\end{eqnarray}
The intensity of the output is related to the intensity of the intracavity field,  $|C_n|^2 = 2\Gamma_n |A_n|^2$.

\begin{figure}[h]
\centering
\includegraphics[width=0.9\columnwidth]{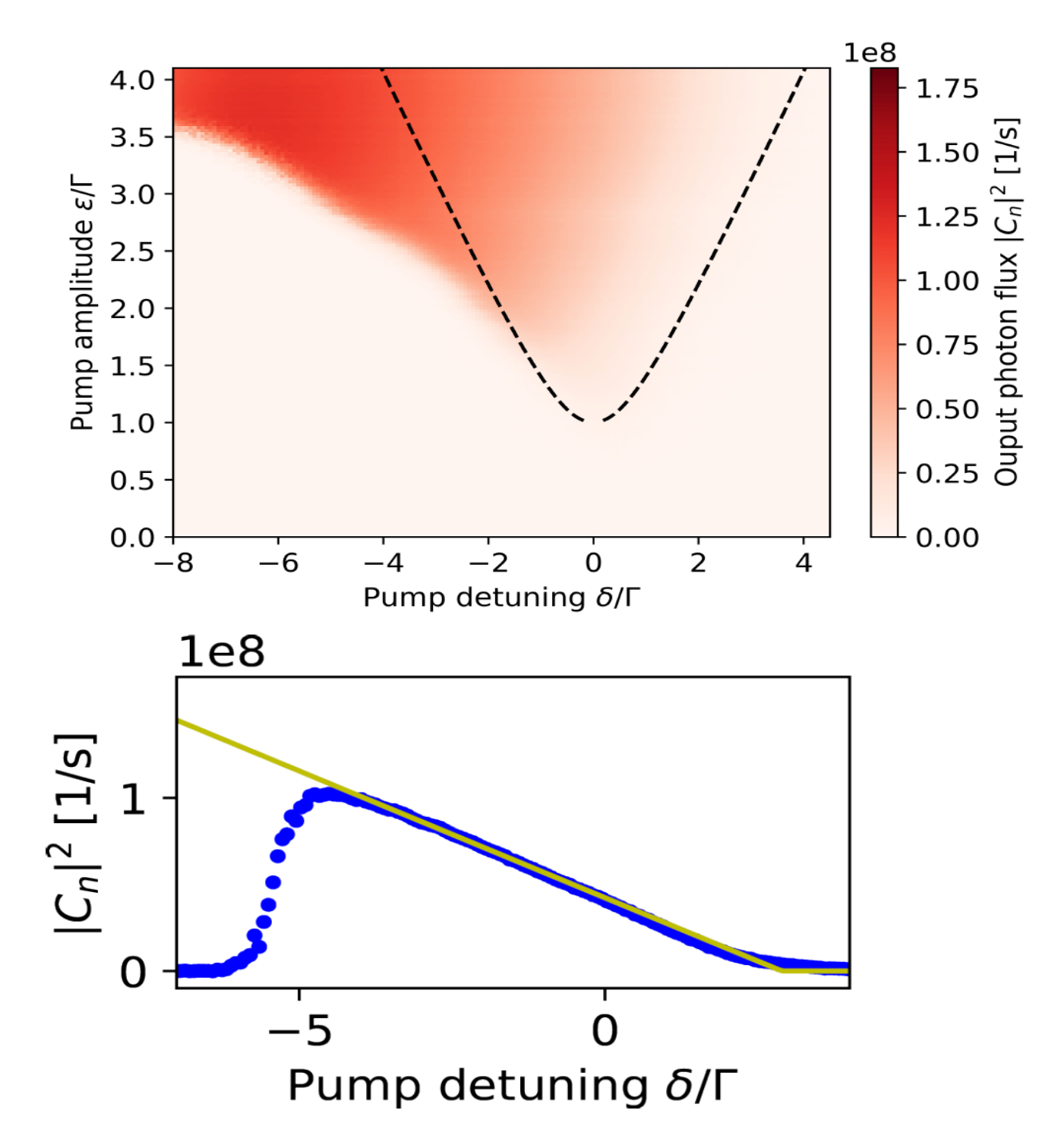}
\caption{Experimentally measured intensity of parametric oscillation as function of pump detuning and intensity\cite{BengtssonPRB2018}. Dashed line in upper panel shows instability threshold; oscillation intensity gradually increases at the threshold at  
$\delta>\delta_{th}$, and sharply disappears at  $\delta<-\delta_{th}$. Lower panel compares a theoretical prediction\cite{WustmannPRB2013} (yellow line) and experimental observation (blue dots).}
\label{fig:Degenerate_exp}
\end{figure}

In Fig.~\ref{Fig:Degenerate_th} the properties of oscillatory states are illustrated. 
The stable oscillatory state is born at the positive threshold branch, $\delta=\delta_{th}$ (blue line in the upper panel), as the result of instability of the ground state. This line corresponds to the second order phase transition. The oscillation amplitude, \Eq{eq:A_osc_deg}, remains finite at any given detuning but grows with increasing red detuning. Under the effect of an input, the degenerate cavity amplitudes described by \Eq{eq:A_osc_deg} split  into two branches (orange lines in central panel). 
 The picture at the  negative threshold branch,  $\delta=-\delta_{th}$, is more complex: here a new  excited state emerges (shown by dashed lines in center panel) but it is unstable. At the same time the stable oscillatory state persists, and  coexists  with the re-established  ground state at increasing red detuning . 
 
 This theoretical picture of coexisting excited and ground states is, however, not verified by the experiment\cite{BengtssonPRB2018} as shown in Fig.~\ref{fig:Degenerate_exp}. The experimental data shows that the excited state undergoes a cross over to the ground state as the red detuning increases.  This crossover resembles a phase transition of the first order,  where noise effects need to be included to characterize the transition. 
Fig.~\ref{Fig:Chris} shows the measured histograms of the cavity output quadratures, at three different values of the pump detuning\cite{WilsonPRL2010}. Such histograms are generated through the sampling of of current-voltage data (which represent two orthogonal quadratures) over a sufficient measurement time, and thus give a picture of the stable cavity states and fluctuations around them. Here we find all three oscillatory regimes: squeezed ground state noise (a), double degenerate oscillatory state (b), and oscillations coexisting with the ground state (right). 

\begin{figure}[h]
\centering
\includegraphics[width=\columnwidth]{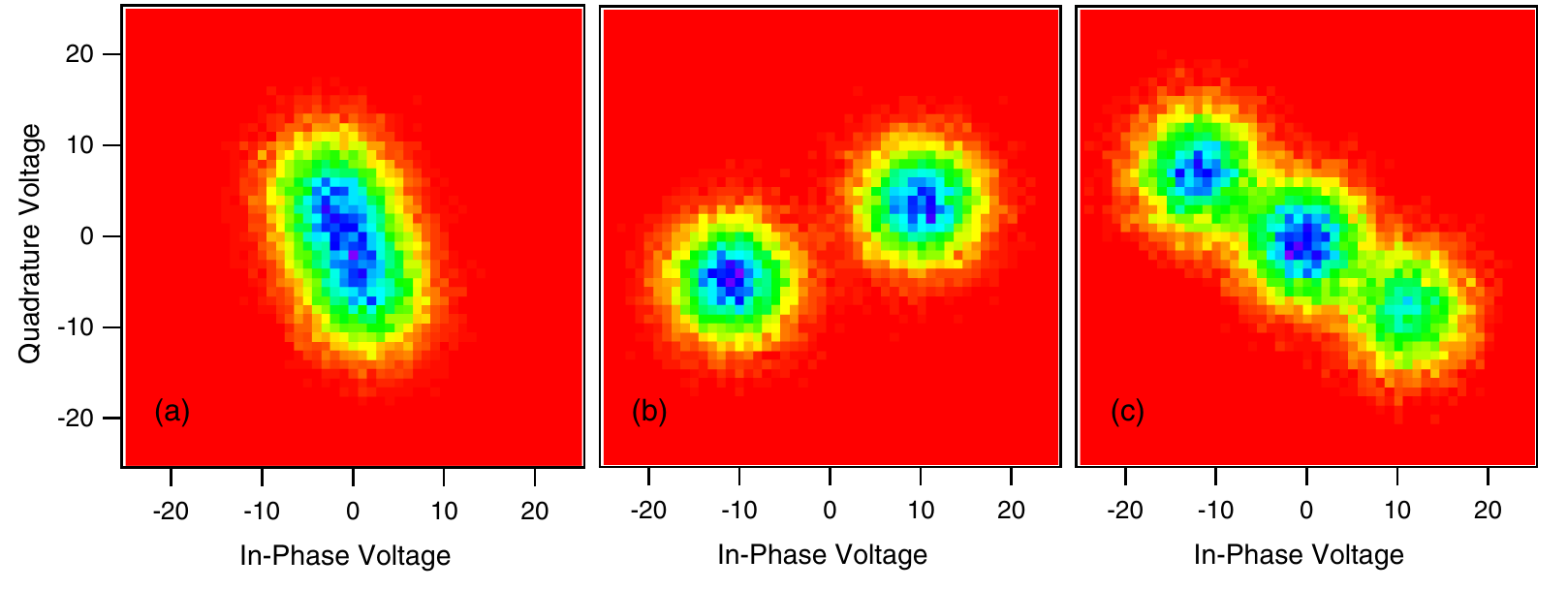}
\caption{Experimental histograms of output radiation at different detuning regions \cite{WilsonPRL2010} corresponding to phase portraits in  
Fig.~\ref{Fig:Degenerate_th}: (a) squeezed vacuum at $\delta>\delta_{th}$, (b) two phase degenerate oscillator states at 
$|\delta|< \delta_{th}$, and (c) coexisting oscillations and ground state at  $\delta< -\delta_{th}$.
}
\label{Fig:Chris}
\end{figure}
%

%%%%%%%%%%%%%%%%%%%%%%%%%%%%%%%%%%%%
\subsubsection{Quantum cat states}
\label{sec:cats}

In spite of cumbersome form of the density matrix of parametric oscillator\cite{KryKhe1996,Drummond2002,MilburnDuty2014} and complex physics of quantum interstate transitions \cite{MarthalerPRA2006,MarthalerPRA2007,DykmanPRL2012}, 
the question about the nature of the quantum ground state in the parametric oscillator has a surprisingly simple answer in the case of undamped cavity\cite{Cochrane1999,Puri2017}. Let us revisit the Hamiltonian for the degenerate resonance, \Eq{eq:Heff_deg}, and present it in a factorized form, assuming detuning $\delta = -\alpha_n/2$,
\begin{equation}\label{eq:H_factorized}
H = - {\hbar \alpha_n\over 2} \left(a^{\dag 2} + {\epsilon_n\over\alpha_n} \right) \left(a^{ 2} + {\epsilon_n\over\alpha_n} 
\right) + {\hbar\epsilon_n^2\over  2 \alpha_n} \;.
\end{equation}
Let us then consider a Glauber coherent state $|\beta\ra$, 
\begin{eqnarray}
|\beta\ra = e^{-|\beta|^2/2}\sum_n {\beta^n \over n!} (a^\dag)^n |0\ra
 \end{eqnarray}
and apply the Hamiltonian to such a state. Remembering that 
the coherent state is an eigenstate of an annihilation operator, $a|\beta\ra = \beta |\beta\ra$, we find that the states,
$|\beta_\pm\ra = |\pm i\sqrt{\epsilon_n/\alpha_n}\ra$, are eigenstates of the Hamiltonian \eqref{eq:H_factorized}. Moreover, the average values of the state amplitudes, $\la \beta_\pm |\, a \,|\beta_\pm\ra = \pm i\sqrt{\epsilon_n/\alpha_n}$, coincide with the earlier found quasiclassical amplitudes in   \Eqs{eq:A_osc_deg}-(\ref{eq:theta_osc_deg}) in the limit 
$\alpha_n \ll \epsilon_n$, $\Gamma_n=0$.
Thus one concludes that the stationary state of an undamped parametric oscillator is a quantum cat state - a superposition of the two coherent states,  
\begin{eqnarray}
|\Psi \rangle = C_1\,  |  i \sqrt{\epsilon_n/\alpha_n}\rangle + C_2 \, | -i \sqrt{\epsilon_n/\alpha_n}\rangle\,.
 \end{eqnarray}
%

%
%%%%%%%%%%%%%%%%%%%%%%%%%%%%%%
\subsubsection{Application to qubit readout}

\begin{figure}[h]
\centering
 \includegraphics[width=0.9\columnwidth]{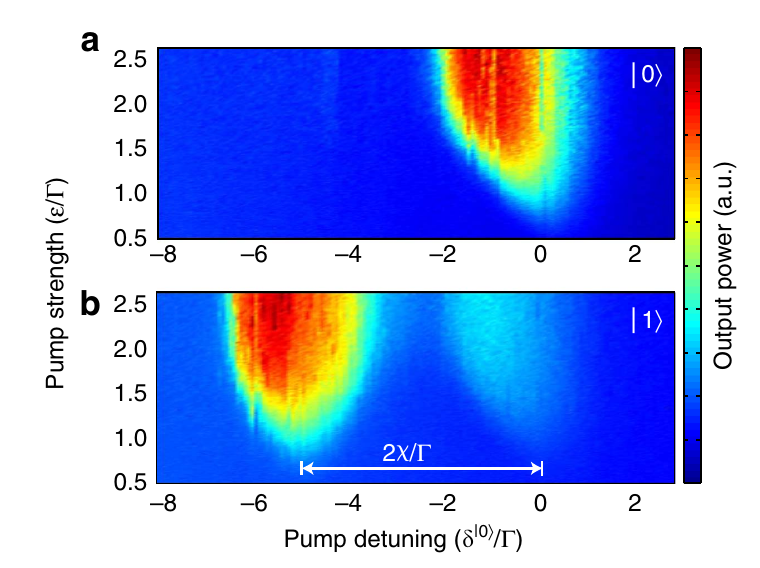}
\caption{Output radiation of parametrically pumped cavity with dispersively coupled qubit, for the ground state (a) and excited state (b) of the qubit, $2\chi$ is a dispersive shift produced by the qubit; the low intensity spot in panel (b) is due to qubit relaxation during measurement\cite{KrantzNatComm2016}.
}
\label{fig:Kranz}
\end{figure}

The existence of parametric oscillation within a well defined interval of detuning can be employed for qubit readout. In c-QED the qubit readout is commonly realized by coupling the  qubit to a linear cavity and 
measuring the cavity resonance\cite{SchoelkopfNat2008}. When the qubit is dispersively coupled to the cavity, it shifts the cavity resonance, and this shift  depends on on the state of the qubit. For a strong qubit-cavity coupling the frequency shift 
exceeds the cavity bandwidth, which enables a single shot readout. The probing signal is usually weak and requires 
subsequent parametric amplification.
The idea of measurement method based on parametric oscillation and realized in\cite{LinNatComm2014,KrantzNatComm2016}  is to combine the readout cavity and parametric amplifier in one device. This is illustrated with experimental data in Fig.~\ref{fig:Kranz}: the panel (a) shows intensity of parametric radiation as function of pump detuning and strength when the qubit is in the ground state, while the panel (b) shows that when the qubit is in the excited state. The distance between the bright regions exceeds the dispersive offset, $2\chi$, exerted by the qubit. The low intensity spot in the panel (b) is due to the qubit relaxation during the measurement. 

%%%%%%%%%%%%%%%%%%%%%%%%%%%%%%%%
\subsection{Nondegenerate oscillations}
\label{sec:ndeg_osc}
Consider now the nondegenerate resonance, and first examine the quasiclassical version of \Eq{eq:EOM_ndeg} using the balanced mode model. In this case, the instability picture is exactly the same as for the degenerate resonance. The instability occurs within the interval, \Eq{eq:threshold}. The self-sustained oscillations appear in both parametrically coupled 
modes having frequencies,  $\omega_j +\delta$, and  equal intensities, $|A_n|^2 = |A_m|^2$. The latter is given by slightly modified \Eq{eq:A_osc_deg}, where the Kerr coefficient is replaced, $\alpha_n \to 3\alpha_n$ because of the cross-Kerr effect. The major difference here is in the phase properties of the oscillations. The sum of the oscillation phases, $\Theta = \theta_n+\theta_m$, is rigorously defined with the same equation as in the degenerate case, $\sin \Theta = \Gamma_n/\epsilon_n$,  however the {\em difference} of the phases, $\psi = \theta_n-\theta_m$ is arbitrary. This implies a continuous degeneracy of the oscillations  with respect to the phase, which is qualitatively different from the discrete (double) phase degeneracy of the degenerate oscillation. 

\begin{figure}[h]
\centering
 \includegraphics[width=\columnwidth]{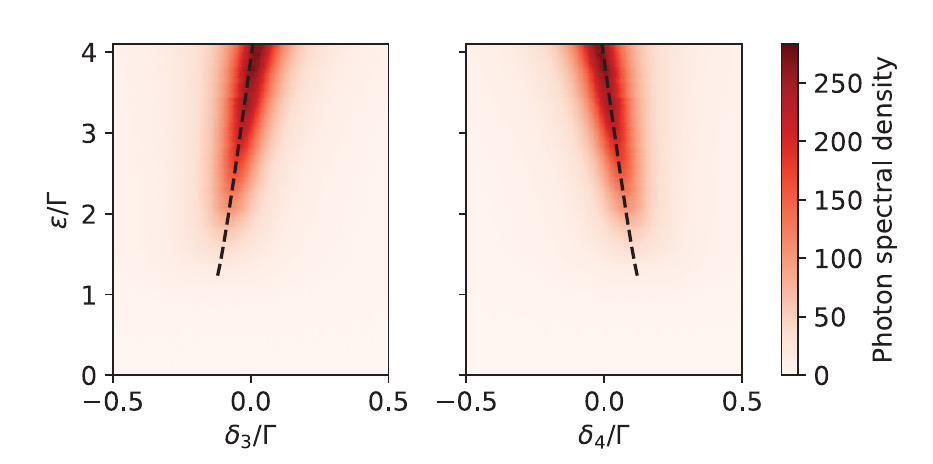}
\caption{Nondegenerate parametric oscillations of modes $n=3$ and $n=4$ observed experimentally \cite{BengtssonPRB2018} emerge at finite detuning from the resonance, $\omega_j + \delta$, and exhibit frequency drift with increasing pumping intensity, in accord with the theory, \Eqs{eq:Delta_thresh} and \eqref{eq:Delta}.
}
\label{Fig:DetuningOscillation}
\end{figure}

In realistic case of different mode parameters, the situation is more complex, as it follows from analysis of  \Eq{eq:nonlinearA}. First, the instability occurs at finite detuning, 
\begin{eqnarray}\label{eq:Delta_thresh}
 \Delta &=& {\Gamma_n  - \Gamma_m  \over \Gamma_n + \Gamma_m}\, \delta \,,
\end{eqnarray}
and the detuning of the emerging oscillation  grows with increasing pumping strength,  
\begin{eqnarray}\label{eq:Delta}
 \Delta(\epsilon_{nm}) = {\Gamma_n \zeta_m - \Gamma_m \zeta_n \over \Gamma_n + \Gamma_m} 
\;.
\end{eqnarray}
These properties of nondegenerate oscillations are confirmed in experiment, as shown in Fig~\ref{Fig:DetuningOscillation}.
Furthermore, the intracavity oscillation intensities  are different, $| A_{n}|^2 / |A_{m}|^2 = \Gamma_m / \Gamma_n$,
however the output intensities are equal, $|C_n|^2 = |C_m|^2$. The latter can be interpreted quantum mechanically as a creation of photons in pairs, giving equal number of photons in each mode. 
The oscillation amplitudes have quantitative difference from the degenerate oscillator,
\begin{eqnarray}\label{eq:An_osc}
|A_{n}|^2 = {2(-\delta \mp \delta_{\rm th}) \Gamma_m\over \alpha_n\Gamma_m + \alpha_m\Gamma_n
+ 2\alpha(\Gamma_n + \Gamma_m)}\,,
\end{eqnarray}
as well as the threshold detuning that defines the ground state instability region,
\begin{eqnarray}\label{eq:delta_thresh}
\delta^2 <  \deltath^2 = \frac{(\Gamma_n+\Gamma_m)^2}{4 \Gamma_n \Gamma_m}\,
 (\epsilon_{nm}^2-\Gamma_n \Gamma_m)
\;.
\end{eqnarray}
However, both quantities have the same dependence on the pumping strength as in the degenerate case. Finally, the equation for the sum of oscillation phases extends in a predictable way the result of \Eq{eq:theta_osc_deg},
\begin{eqnarray}\label{eq:Theta_osc}
\sin\Theta= {\sqrt{\Gamma_n\Gamma_m}\over \epsilon_{nm}}, \quad  \Theta   \in ({\pi\over 2}, \pi) \,. 
\end{eqnarray}
\begin{figure}[h]
\centering
 \includegraphics[width=0.8\columnwidth]{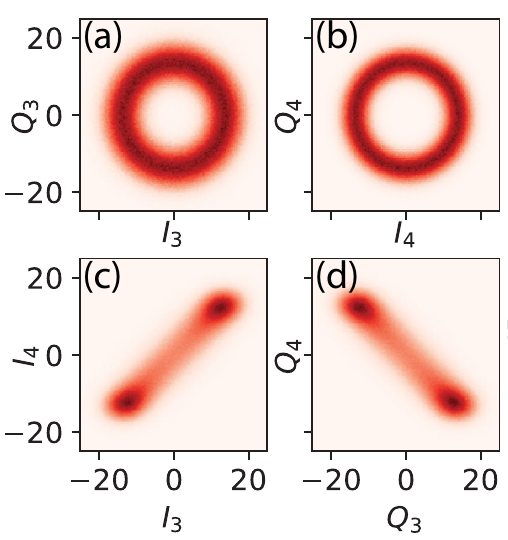}
\caption{Measured quadrature histograms of non-degenerate parametric oscillations in modes $n=3$ and $n=4$\cite{BengtssonPRB2018}.  Panels (a-b): oscillation phases in both modes are evenly spread under effect of noise due to a phase diffusion. Panels (c-d): the quadratures of different modes are anti-correlated,  confirming the preservation of the sum of the mode phases, \Eq{eq:Theta_osc}}
\label{Fig:Donuts}
\end{figure}

In Fig.~\ref{Fig:Donuts}(a-b) experimental histograms in quadrature space are presented for output intensities in both oscillating modes\cite{BengtssonPRB2018}. The histograms  have donut shapes that indicate continuous phase degeneracy. In absence of environmental noise, which is assumed in our calculation, the phases would have specific values defined by initial conditions. In practice, the presence of noise induces a phase diffusion that spreads the phases homogeneously in the steady state. This behavior is common for generators with continuous phase degeneracy (e.g. Van der Pol oscillator \cite{RytovBook}, nondegenerate optical parametric oscillator \cite{ReynaudJMO1991}).  Panels (c-d) show cross correlations between mode quadratures demonstrating the rigid constraint on the sum of the phases imposed by \Eq{eq:Theta_osc}.   
%

%%%%%%%%%%%%%%%%%%%%%%
\subsubsection{Phase locking } 
\label{Phase locking}
Continuous phase degeneracy of non-degenerate parametric oscillations and related phase diffusion leads to considerable broadening of the output  linewidth. This effect is known in lasers  and  microwave generators, where it is eliminated, in particular, by injecting a weak on-resonance signal. \cite{Adler1946} To illustrate the mechanism of the injection phase locking let us  consider the degenerate parametric resonance and add a driving term, $\sqrt{2\Gamma_n} |B_n|(e^{i\theta_B} A_n^\ast + e^{-i\theta_B} A_n)$, to the quasiclassical metapotential, \Eq{eq:Heff_deg_pq}. Proceeding to polar coordinates, $Q_n + i P_n = R_ne^{i\theta_n}$, we get (dropping mode index),
\begin{eqnarray}\label{}
 H(R, \theta) = &-& \frac{\delta}{2}R^2 -  \frac{\alpha}{8}R^4 - {\epsilon} R^2\cos 2\theta \no
&+& 2\sqrt{\Gamma} |B|R\cos(\theta-\theta_B)\,.
\end{eqnarray}
Then we see that the unperturbed metapotential obeys the symmetry $\theta \to \theta+\pi$, and it is violated by the driving term. This introduces asymmetry in the metapotential, as illustrated in Fig.~\ref{Fig:tilted_meta}, which is also seen in the phase portraits in Fig.~\ref{Fig:Degenerate_th} and  manifested by the splitted response line in this figure.  Due to the  
asymmetry, one of the formerly degenerate steady states will be lower in energy than the other and thus will be populated with higher probability. If the asymmetry is made sufficiently strong, only this state will persist.
The phase locking effect was employed for qubit readout in the parametric oscillation regime\cite{LinNatComm2014}.    
\begin{figure}[h]
\centering
 \includegraphics[width=0.5\columnwidth]{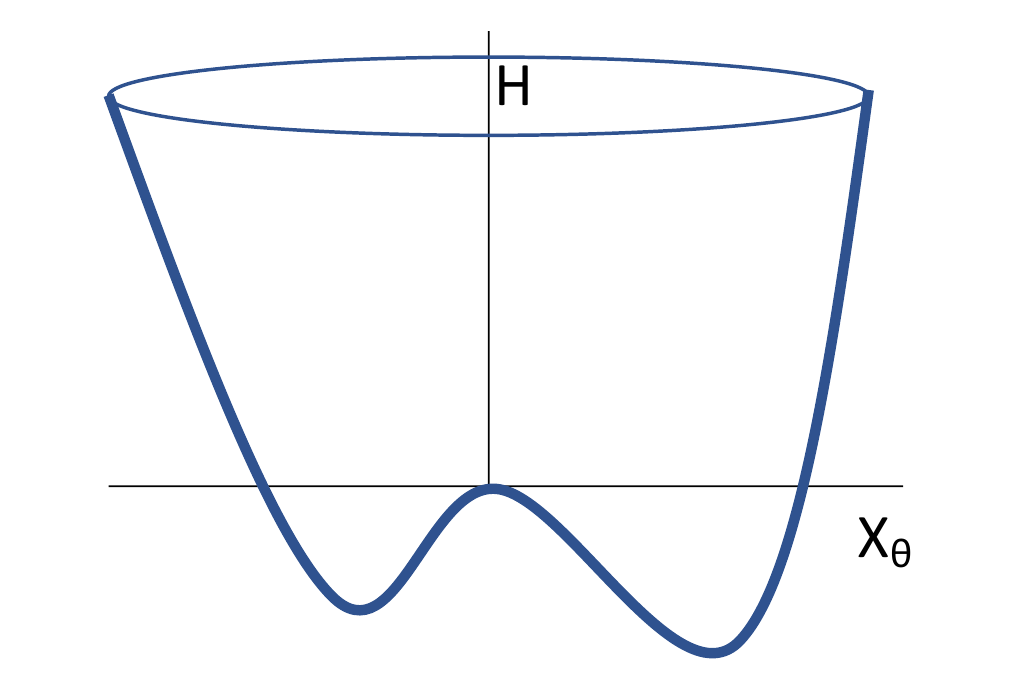}
\caption{Tilted metapotential of the degenerate oscillator, cut along optimum phase direction.}
\label{Fig:tilted_meta}
\end{figure}

Similar argument applies to the nondegenerate oscillator, the tilt of the metapotential produced by the driving term removes the phase degeneracy. The value of the locked phase is determined by the phase of the input, the corresponding relation was found in \cite{Wustmann2017} for balanced mode model,
\begin{eqnarray} \label{eq_psi}
\theta_n =  \theta_B - {\rm arctan}\,(\Gamma/ \zeta_-) \,.
\end{eqnarray}
The phase locking effect was observed experimentally \cite{BengtssonPRB2018}, the resulting quadrature histograms are presented 
in Fig.~\ref{Fig:Locking_Andreas}(a-b). Here the spread of the phase in both modes (shown for mode 3) diminishes with 
increasing strength of the locking input.
Correspondingly, the radiation line width dramatically decreases, by several orders of magnitude for few photon input, panel (c).

\begin{figure}[h]
\centering
 \includegraphics[width=\columnwidth]{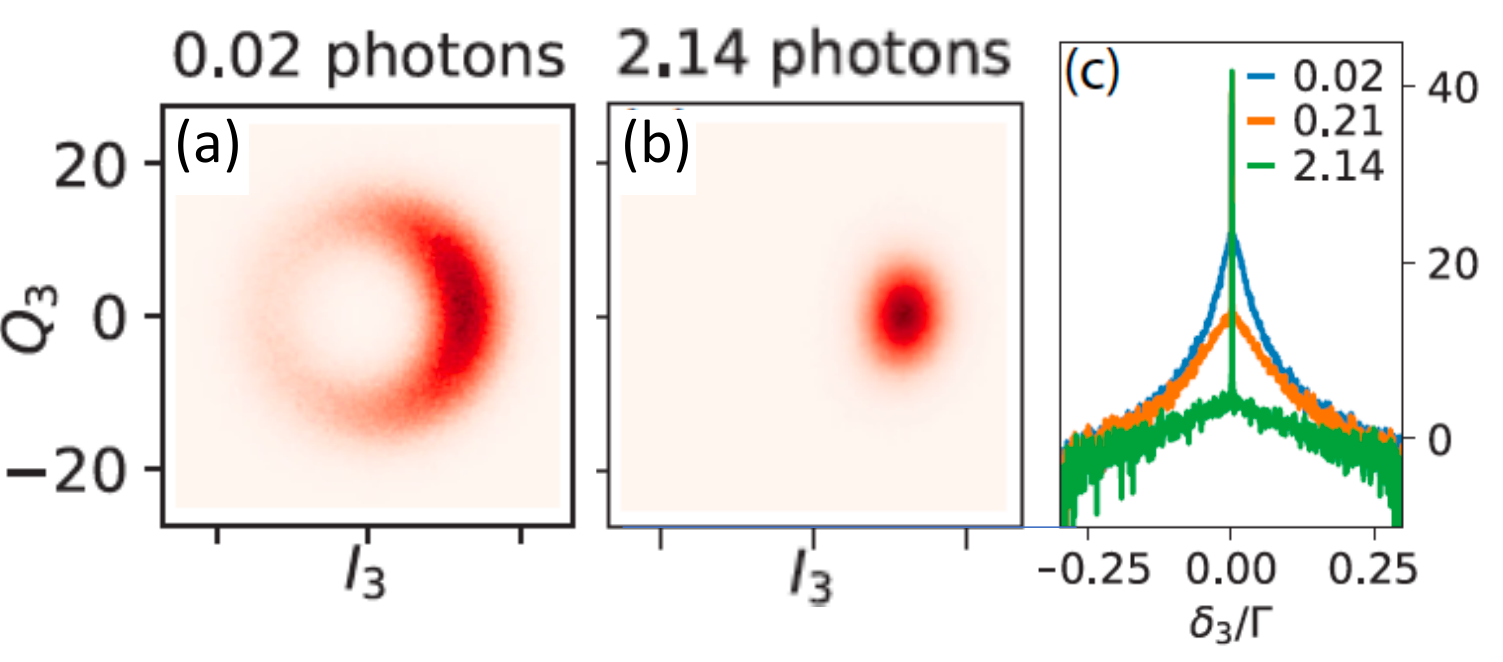}
\caption{Phase locking effect under injection of on-resonance signal\cite{BengtssonPRB2018}.  (a-b) Histograms for output of mode $n=3$ at different injected photon numbers, indicating suppression of phase diffusion under increasing injection. (c)  Measured output photon spectral density (in dB) at different injected photon injection numbers, the linewidth decreases by orders of magnitude with increasing number of injected photons.
}
\label{Fig:Locking_Andreas}
\end{figure}
%

%%%%%%%%%%%%%%%%%%%%%%%%%%%%%%%%%%%%%
\subsection{Subharmonic oscillations}
\label{sec:subharmonics}

In this section we consider a different class of oscillatory states - subharmonic oscillations\cite{JordanBook,Hayashi} that are associated with temporal modulation of the nonlinearity coefficients\cite{GuoPRL2013,SvenssonAPL2018}. As it was mentioned before, the modulation of the  magnetic flux 
through the SQUID affects all the harmonics of the Josephson inductance.  Here we will see that driving the flux with frequency close to multiples of a mode frequency, $\Omega\approx N\omega_n$ will excite oscillation of this mode. 

In Fig.~\ref{Fig:3_4_histograms} we present the histograms of the cavity output, when it is parametrically driven with frequencies close to multiples of the fundamental cavity mode, $3\omega_0$, $4\omega_0$ and $5\omega_0$  \cite{SvenssonAPL2018}.  The outputs presented in right panels exhibit multiple bright spots indicating multiple phase degeneracy of the
 oscillatory states. At the left panels these states appear to coexist with the ground state of the cavity (cf.  Fig.~\ref{Fig:Chris} for the degenerate parametric oscillator).  

\begin{figure}[h]
\centering
 \includegraphics[width=0.9\columnwidth]{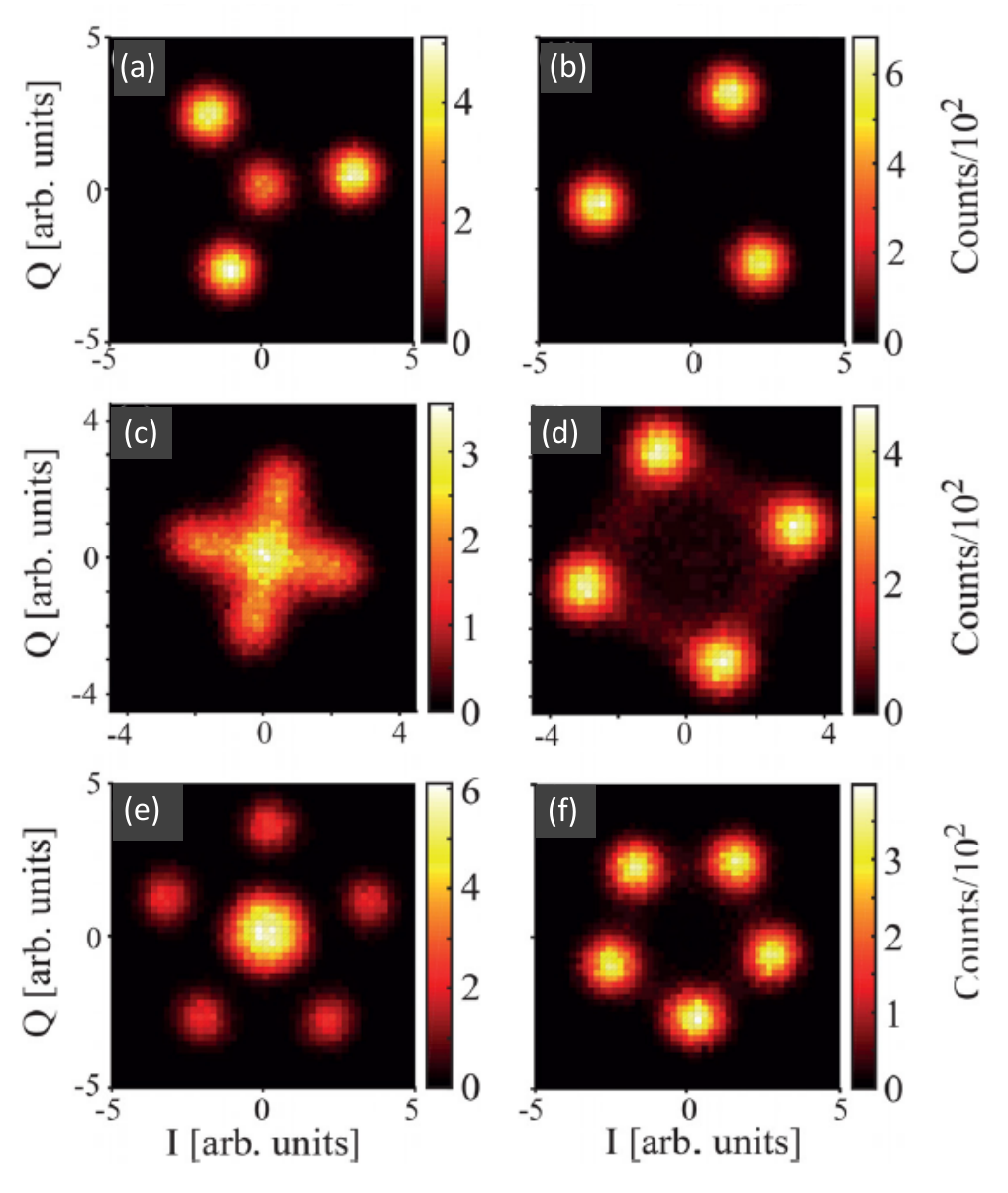}
\caption{ Experimentally measured histograms of N=3 (a,b), N=4 (c,d)  and N=5 (e,f) subharmonic oscillations demonstrate N-fold phase degeneracy of oscillator states \cite{SvenssonLicense}. Central bright spot at left panels indicates cavity ground state, other bright spots refer to oscillator excited states; histograms in the left panels are measured at the edge of the oscillation visibility demonstrating cross over  to the ground state; histograms in the right panels are measured in the region of intense output. Blurred lines on panels (c)-(d) reveal interstate transitions.  
}
\label{Fig:3_4_histograms}
\end{figure}

To understand the origin of these subharmonic oscillatory states we revisit \Eq{eq:V} for the Josephson potential and retain in the driving term, $\propto \delta f$, higher order terms with respect to the phase, $\propto \phi^{n}(d,t)$. Then one finds that for the driving frequency $\Omega\approx N\omega_0$ the $N$-th order term of the expansion is resonant. At the same time, the lowest order Kerr term in the static part of the potential remains dominant. As a result, the quasiclassical Langevin equation takes the form (omitting mode index),
\begin{eqnarray}\label{eq:EOM_nomega}
i\dot A + (\delta  + i\Gamma + \alpha |A|^2) A + \epsilon_{(N)} (A^{\ast})^{ N-1} = 0 . 
\end{eqnarray}
The  pumping coefficient here has a general form,  
\begin{eqnarray}
\hbar\epsilon_{(N)} = c_{N}  {\delta f\over 2} E_J \sin{F\over 2}\, \, s_0^{N} , 
\end{eqnarray}
where $c_{N}$ is a numerical coefficient,   the Kerr coefficient  $\alpha$  is the same as in \Eq{eq:def_alpha_n}. 

Introducing the amplitude and the phase of the oscillation, $A = |A|e^{i\theta}$, we find stationary values of the phase, 
\begin{eqnarray}
\theta = \theta_0 + {2\pi k\over N}, 
\; k = 1,\ldots  N,  \;\; \sin(N\theta_0) = {\Gamma\over \epsilon_{(N)} |A|^{N-2}}.  \no 
\end{eqnarray}
Thus the solution has a discrete, $N$-fold phase degeneracy, which is a general property of subharmonic oscillations.
This is illustrated in  Fig.~\ref{Fig:3portrait} with a phase portrait for the lowest order subharmonic oscillation, $N=3$. There are three stationary excited states with equal amplitude absolute values that are phase shifted by $2\pi/3$. 
\begin{figure}[h]
\centering
 \includegraphics[width=0.7\columnwidth]{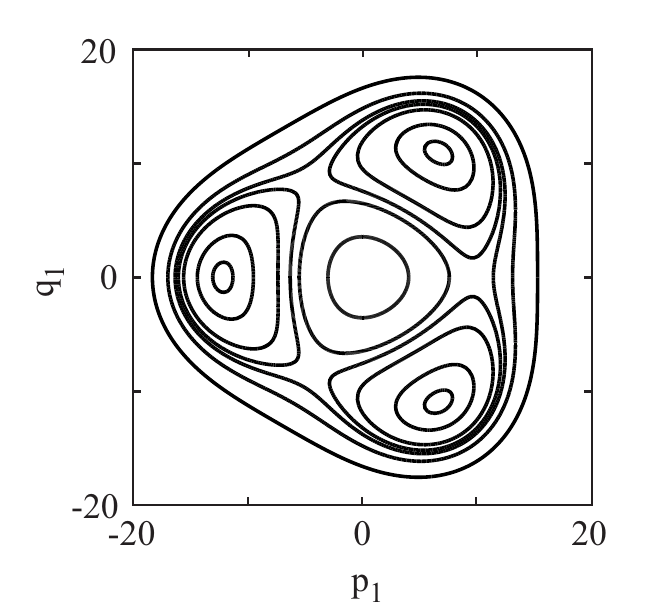}
\caption{Phase portrait of the third order subharmonic oscillation, the period tripling\cite{SvenssonPRB2017}.The oscillator has four steady states - the ground state at the origin, and  three excited states at equal distance from the origin having phases shifted by $2\pi/3$; all these states are stable. }
\label{Fig:3portrait}
\end{figure}

As the expansion of $\cos\phi(d,t)$ in \Eq{eq:V}
produces only even orders of $\phi$, the above derivation only holds for even subharmonics, $N=2k$. To excite the odd subharmonics, $N=2k+1$,  one should employ an asymmetric SQUID with different Josephson energies of the junctions, $E_{J1} - E_{J2} = 2E_- \neq 0$ \cite{SvenssonAPL2018}. In this case, the potential in \Eq{eq:V} 
acquires an additional term proportional to the SQUID asymmetry, $- 2E_- \sin[f(t)/2]\sin\phi(d,t)$, which contains the odd-order terms.  Linearization over $\delta f(t)$ yields the driving term, $- \delta f(t) (E_- /\cos(F/2))\,\sin(\phi(d,t) - \phi_0)$, that enables excitation of the odd subharmonic oscillations  being driven with the odd multiples of the mode frequency (here $\phi_0\propto E_-$ is a static phase shift).   The pumping coefficient in this case has the form,
$\hbar\epsilon_{N} = c_N (\delta f/2)\, E_-   \cos(F/2) s_0^N$ (for small asymmetry), which has similar scaling with growing $N$ as the even subharmonics. 

In the case of odd subharmonics, however, an additional effect occurs: The driving term here contains  a linear component $\propto \phi(d,t)$, which results in injection of  pumping field directly into the cavity. The amplitude of this field can be appreciable if the pumping frequency is close to one of the cavity resonances.  This intracavity field produces an additional pumping effect, current pumping, similar to the degenerate parametric resonance discussed in Sec.~\ref{sec:strong_weak} for amplification of weak signal in presence of strong field. 
The effective pumping strength then gets an addition $\propto s_0^{N-1}s_m A_m$, where $A_m$ is the amplitude of excited cavity mode. This intracavity field also produces a cross Kerr effect, $\propto |A_m|^2$.

\begin{figure}[h]
\centering
\includegraphics[width=\columnwidth]{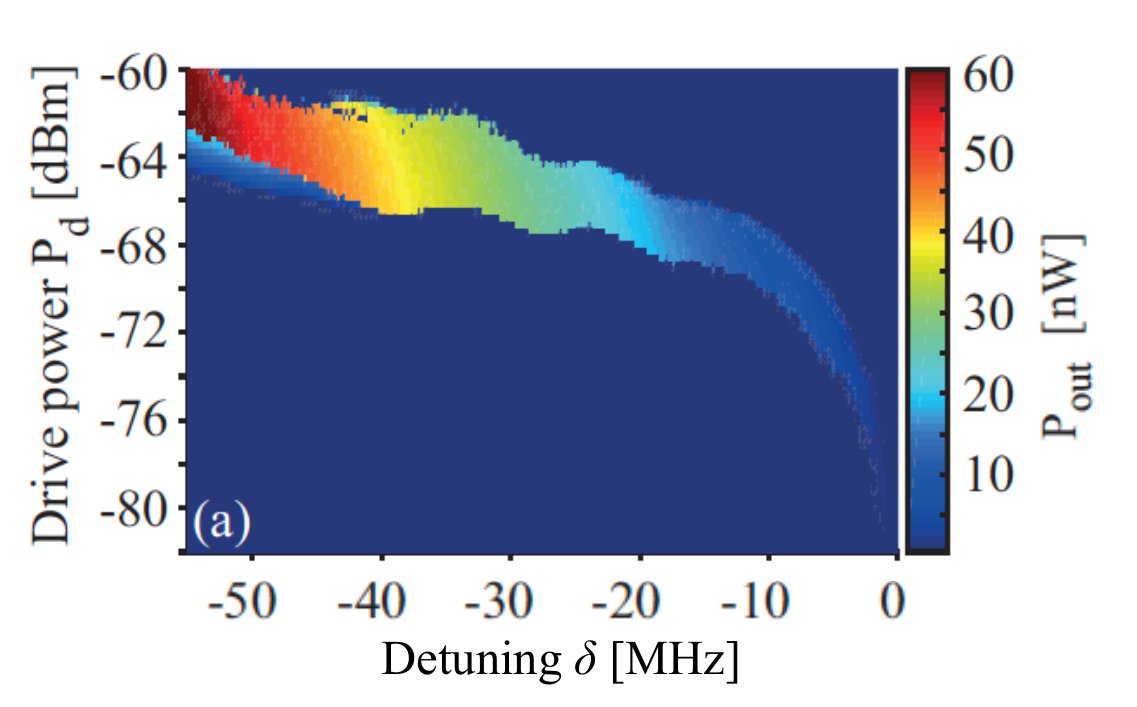}
\caption{Experimentally measured output intensity of N=3 oscillation as function of detuning and driving power; the oscillation spreads far outside the cavity bandwidth (< 200 KHz) towards the red detuning, oscillation intensity increases with increasing detuning. (Adopted from\cite{SvenssonPhD}, courtesy of I.-M. Svensson.) }
\label{Fig:3omega}
\end{figure}

The period tripling oscillation was experimentally studied in detail in Ref.~\cite{SvenssonPRB2017}.  
 The ntensity of the observed output is presented in Fig.~\ref{Fig:3omega}. Appreciable output signal is detected within a rather narrow strip of pumping intensity but vast region of negative detuning, starting from near exact resonance and spreading  far beyond the cavity bandwidth ($\Gamma_0/2\pi$ = 190 KHz).
Moreover, the output intensity  is found to grow with the detuning. This is clearly due to the Kerr effect, which shifts the cavity resonance proportionally to the field intensity. 
To achieve a better control over the device parameters and enable quantitative comparison with the theory,  the current pumping scheme was employed.
The flux pumping was disabled and  instead a strong calibrated signal was injected into the cavity with frequency  
$\Omega= 3(\omega_0+\delta)$, and the response was measured at $\omega_0+\delta$.  Comparison with theory gives a good agreement for the frequency threshold and the upper boundary of of the oscillation region, which coincides 
 with the theoretical boundary of the oscillation existence\cite{SvenssonPRB2017}. However, the observed lower boundary lies far above the theoretical prediction. This can be understood as an indication of the first order phase transition between the oscillator excited states and the ground state, which is analogous to the behaviour of the parametric oscillations at red detuning discussed in Sec.~\ref{sec:oscillation_deg} and shown in Fig.~\ref {fig:Degenerate_exp}. An important difference of the period tripling oscillation from the parametric oscillation is that they do not emerge as the result of  instability of the ground state - the latter always maintains  stability. The oscillation amplitude jumps to a finite value at the edge of existence and maintains stability within the whole region of existence.

%%%%%%%%%%%%%%%%%%%%%%%%%%%%%%%%%%
%%%%%%%%%%%%%%%%%%%%%%%%%%%%%%%
\section{Quantum fluctuations}
\label{sec:quantum}
So far we studied parametric phenomena in the classical domain. Now we proceed to discussion of quantum properties of output radiation: quantum statistics and correlation functions of outgoing photons. We limit our discussion to small quantum fluctuations around classical steady states. These fluctuations are conveniently described with linearized Langevin equations. Application of Langevin equations to large quantum fluctuations, like  critical fluctuations near the instability threshold or quantum jumps within multistability regions, 
requires solving nonlinear operator equations. To circumvent this difficulty alternative methods are  applied.  Some exact results on critical fluctuations were obtained with a  master equation approach ~\cite{KryKhe1996,Drummond2002, MilburnDuty2014}, quantum transitions among degenerate oscillatory states were investigated in Refs.~\cite{MarthalerPRA2006,MarthalerPRA2007,DykmanPRL2012,GuoPRL2013, AnkerholdPRA2017}. 

From the physics viewpoint the noise of the output radiation results from the environmental noise that enters the input port and is "processed" by the parametrically driven cavity. For equilibrium environment the input noise consists of the classical thermal component and the quantum noise. In what follows we restrict to the zero temperature of the environment and consider only quantum noise originated from the vacuum fluctuations.  

%%%%%%%%%%%%%%%%%%%%%%%%%%%%
\subsection{Squeezed vacuum}
\subsubsection{Two-mode  entanglement}
In the linear amplification regime, the output noise consists of coupled signal and idler modes, \Eq{eq:BT}. In the quantum regime the photons of these modes are strongly correlated. The quantum properties of output noise are fully described with the quantum BT, \Eq{eq:BT} where quantities $c_j(\Delta)$ are bosonic operators satisfying commutation relations, 
$[c_j(\Delta), c^\dag_{j'}(\Delta')] = \delta_{jj'}\delta(\Delta - \Delta')$. The $uv$-coefficients are given by \Eqs{eq:u_ndeg_linearized}-\eqref{eq:v_ndeg_linearized}. The correlation of two modes is also relevant for the nonlinear amplification under the degenerate resonance, when a classical field from either input signal or parametric oscillation fills the cavity. The results obtained below are extended to this case simply by including the  corresponding cross-Kerr effect, $\delta\to \zeta_n$, \Eq{zeta_ndeg}. 

Consider a single quantum noise mode associated with detuning $\Delta$. The input vacuum state is defined by relation $b_n(\Delta)|0\ra =0$. The output state for such an input is not a vacuum state, $ c_n(\Delta) |0\ra \neq 0$. To evaluate the output vacuum state we consider a unitary transformation that generates the BT\cite{SqueezingBook},
\begin{eqnarray}\label{eq:TMSoperator_ndeg}
&& c_n(\Delta) = S\, b_n(\Delta)\, S^\dag \,,  \\    
&& S = \exp\left[ \int_{-\infty}^\infty d\Delta' \xi^\ast(\Delta') b_n (\Delta') b_m (-\Delta') - \text{H.c.}\right]
\,.\nonumber
\end{eqnarray}
It is straightforward to check that this transformation reproduces \Eq{eq:BT}, up to a phase factor, if 
$\xi(\Delta) = r(\Delta)e^{i\rho(\Delta)}$, where $\rho(\Delta) = {\rm arg}\, (v_n(\Delta)/u_n(\Delta))$ is  mode independent. The operator, \eqref{eq:TMSoperator_ndeg} is the squeezing operator that  transforms the input vacuum into the squeezed output vacuum $|0'\ra$ \cite{Perelomov1977,Collet1988}
\begin{eqnarray}\label{eq:squeezedvac_ndeg}
 |0' \rangle &=& S|0\ra \no
&=&  \sum_{N=0}^{\infty} {g^N(\Delta) \over \cosh r(\Delta)}
 \, | N(n,\Delta)\rangle\, | N(m, -\Delta) \rangle\,,
\end{eqnarray}
where 
$g(\Delta) = -\tanh r(\Delta) e^{i \rho(\Delta)} = -v_n(\Delta)/u_n(\Delta)$, 
and  $|N(n,\Delta)\rangle$ is the N-photon state at frequency $\Delta$ in corresponding rotating frame. 
The squeezed vacuum consists therefore of the superposition of multiphoton states, each consisting of equal number of photons of modes $(n,\Delta)$ and $(m,-\Delta)$. These modes are therefore entangled, while states with different $|\Delta|$ are uncorrelated. 

The squeezing operator for the degenerate resonance has similar form as \Eq{eq:TMSoperator_ndeg}, where $m=n$
and the lower integration limit is set to zero.

The squeezed vacuum has close resemblance to the BCS ground state in the theory of superconductivity (cf. Ref.~\cite{SaraivaPRL2017}). Indeed, consider the limit of small squeezing parameter, $r\ll1$, then \Eq{eq:squeezedvac_ndeg} is an expansion over this small parameter, and in lowest order it reduces to,
\begin{eqnarray}\label{}
 |0' \rangle &\propto &  (  u(\Delta) -  v(\Delta) b_n^\dag (\Delta)b_m^\dag (-\Delta) \, | 0 \rangle\,.
\end{eqnarray}
This is the exact analog of the BCS ground state \cite{BCS}, where the fermionic operators are replaced with bosonic ones, and the correlated states, $(p ,\sigma)$ and$(-p , -\sigma)$, in the momentum-spin space of a superconductor are replaced with the correlated modes, $(n,\Delta)$ and $(m,-\Delta)$. 

It is interesting that complex correlated structure of the output noise cannot be observed by measuring a  single mode. In such a measurement one gets access only to the reduced density matrix of the mode, which is obtained by averaging the two-mode density matrix of the squeezed vacuum over the second mode,
\begin{eqnarray}\label{reducedDM}
\rho_n &=& {\rm Tr}_m  |0' \ra \la0'| \no
&=&  \sum_{N=0}^\infty {[\tanh^2 r(\Delta)]^N \over \cosh^2 r(\Delta)} |N(n,\Delta) \ra\la N(n,\Delta)| .
\end{eqnarray}
This equation describes a photon thermal state with an effective temperature,
\begin{eqnarray}\label{Teff}
kT^\ast(\Delta) = {\hbar \Delta\over \ln\tanh^2 r(\Delta)}.  
\end{eqnarray}

The degree of the photon-photon correlation is quantified with the entanglement entropy\cite{Entropy}, 
$E[\rho_n] = - {\rm Tr}\,(\rho_n \ln \rho_n)$. Such a function is equal to zero for a pure product state. For the squeezed vacuum,
\begin{eqnarray}\label{E_entangl}
&&E[\rho_n]  =   \\
&& \cosh^2r(\Delta) \ln\cosh^2r(\Delta) 
  - \sinh^2 r(\Delta) \ln \sinh^2 r(\Delta) . \nonumber
\end{eqnarray}
At the zero squeezing parameter the entanglement entropy vanishes, and it grows with increasing squeezing parameter, $E[\rho_n] \sim 2r(\Delta)$ at $r\gg 1$.  The maximum value of the entanglement can be estimated using  \Eqs{Gnonlinear}-\eqref{eq:A1approx@threshold} for the maximum gain at the parametric threshold. Assuming for the vacuum state, $|B|^2/\Gamma\sim 1$, and bearing in mind the relation, $2r \approx  \ln G$, we find\cite{WustmannPRB2013}, 
\begin{eqnarray}\label{E_entangl}
{\rm max}\,E[\rho_n]  \approx {2\over 3} \ln {\Gamma_n\over \alpha_n}
   \end{eqnarray}
All the essential parametric cavity characteristics are quantified with the squeezing parameter:  gain, squeezing, entanglement, effective vacuum temperature, and they have maximum values defined by the ratio of the dissipation over the nonlinearity.

%%%%%%%%%%%%%%%%%%%%%%%%%%%%%%
\subsubsection{ Four-mode entanglement}
The results of the previous section do not directly apply to the noise in presence of a strong signal under non-degenerate resonance. It is because now four modes become coupled. To find the form of the squeezing operator in this case we resort to the balanced mode model and BT in the supermode basis, \Eqs{BTsigma}-\eqref{eq:uBT_pmbasis}. These equations have the form of the equations for the degenerate resonance, and therefore they can be written on the form, $c_\sigma = Sb_\sigma S^\dag$, with the squeezing operator being the product of operators in \Eq{eq:TMSoperator_ndeg} for both supermodes, 
\begin{eqnarray}\label{S_supermode}
S = e^{  \sum_\sigma \int_0^\infty d\Delta' \xi_\sigma^\ast(\Delta') b_\sigma(\Delta')  
b_\sigma(-\Delta')  - {\rm H.c.}}  \,, 
\end{eqnarray}
here $\xi_\sigma(\Delta)= r_\sigma(\Delta)e^{i\rho_\sigma(\Delta)}$, 
$\rho_\sigma(\Delta) = {\rm arg}(\,v_\sigma(\Delta) / u _\sigma(\Delta))$. 

To get equation for the squeezed vacuum in the original basis, we note that $S$ is a scalar in the supermode space, therefore it is not affected under rotation to the original basis. Furthermore, presenting the exponent in \Eq{S_supermode} in the original basis, we get\cite{Wustmann2017},
\begin{eqnarray}\label{Psiabove}
&& |0'\rangle  =   {1 \over \cosh r_+(\Delta)\cosh r_-(\Delta)} \nonumber\\
&& \times  \exp\left[{g_+ + g_- \over2} \left( e^{i\psi}b^\dag_n(\Delta)b^\dag_n(-\Delta)  
+ e^{-i\psi}b^\dag_m(\Delta)b^\dag_m(-\Delta)\right) \right] \nonumber\\
&&\times \exp\left[{g_+ - g_- \over2} \left(b^\dag_n(\Delta)b^\dag_m(-\Delta) + b^\dag_m(\Delta)b^\dag_n(-\Delta)\right)\right] |0\rangle\,, \nonumber
\end{eqnarray}
where $g_\sigma = -\tanh r_\sigma  e^{i \rho_\sigma}$.
This four-mode squeezed vacuum is a superposition of multiphoton states that contain all possible pairwise combinations from the quartet, $(1,\pm\Delta)$, $(2,\pm\Delta)$. 
It is worth noting that the admixture of the pairs from the same mode (second line in the equation) is entirely defined by the intracavity field, $A_n$, while the coefficient $g_+ + g_-$ turns to zero when $A_n\to 0$. Furthermore, this contribution is  sensitive to the phase difference $\psi$ of the strong field modes.  All the properties of the four-mode squeezed output can be evaluated using the supermode basis.

%%%%%%%%%%%%%%%%%%%%%%%%%%%%%
\subsection{Homodyne detection and SNR}
\label{sec:homodyne}
In circuit-QED the output field is measured by measuring voltage at the output port. The voltage is related to the phase of the output field via the Josephson relation, $V(t) = (\hbar/2e)\dot\phi(t,0)$, hence it represents one of the quadrature. The measurement is usually done by using a homodyne detection scheme: the output field is mixed with a strong field of a local oscillator (LO),
 $A_{LO}e^{i(\omega_n+\delta)t + i\theta} + {\rm c.c.}$, and a low frequency envelope is filtered out producing a quadrature that depends on the phase of the local oscillator\cite{WalMil2008},
\begin{eqnarray}\label{eq:quadrature}
X^\theta(t) &=&  [(C_{n}(t) +  c_{n}(t) ) \,e^{-i\theta} + {\rm h.c.}] \, .
\end{eqnarray}
The spectrum of this output is concentrated around the  frequencies of the measured mode, $\omega_n+\delta$, within  bandwidths, $\Gamma_{n}$. In \Eq{eq:quadrature} $C_n(t)$ is the classical component of the output and $c_n(t)$ is the noise component represented by the bosonic operator.  Variation of the local oscillator phase $\theta$ allows to explore all the output quadratures.  

The output is quantified with a spectral power defined as,\cite{RytovBook,Yurke_DruFic2004}
\begin{eqnarray}\label{eq:spectralpower}
P(\Delta) = \lim_{T\to\infty} \,{1\over 2T}\left| \int _{-T}^{T} dt \,X(t)  e^{i\Delta t} \right|^2  = P_0(\Delta) + S(\Delta) \,,\nonumber\\
\end{eqnarray}
where $P_0(\Delta)$ represents the classical component and has the form for the on-resonance input tone,  
$\Delta=0$, 
\begin{eqnarray}\label{eq:P0}
P_0(\Delta) = 2\pi\delta(\Delta) \left(C_n e^{-i\theta} + C_n^\ast e^{i\theta}\right)^2 . 
\end{eqnarray}
The second term in \Eq{eq:spectralpower} describes the noise, and it is commonly quantified with the squeezing spectral density,
\begin{eqnarray}\label{eq:spectraldensity}
S_n^\theta(\Delta) = \int_{-\infty}^{\infty} dt \, e^{i\Delta t}\la x_n^\theta(t)x_n^\theta(0)\ra \,.
\end{eqnarray}
The quantum expectation values here are evaluated with respect to the input vacuum state.
This spectral density can be expressed through the Fourier harmonics of the noise quadratures
\begin{eqnarray}\label{eq:spectral_x}
 x_n^\theta(\Delta) &=&  
\int_{-\infty}^{\infty} {dt\over \sqrt{2\pi}}\,  e^{i\Delta t}  x^\theta_n (t)  \no
&=& c_{n}(\Delta)   e^{-i\theta} + c_{n}^\dag(-\Delta)  e^{i\theta} \,,
\end{eqnarray}
giving
\begin{eqnarray}\label{eq:spectraldensit_2}
 S_{n}^{\theta}(\Delta) &=& \int _{-\infty}^{\infty}  d\Delta'
\,\langle x_n^{\theta}(\Delta) x_n^{\theta}(\Delta')\rangle \,.
\end{eqnarray}

Here we present the noise spectral densities for different amplification regimes and evaluate the signal to noise ratio (SNR) of the output that certifies the amplification quality.  The SNR is defined,
\begin{eqnarray}\label{eq:SNR_def}
\text{SNR} = {\overline{P_0^\theta(\Delta)}\over \overline{S_n^\theta(\Delta)}},
\end{eqnarray}
where integration is made over some bandwidth, $(-\bar\Delta/2, \bar\Delta /2)$. 
For the input, the signal power is, $\overline{P_0^\theta} = 8\pi |B_n|^2 \cos^2(\theta-\theta_B)$, and 
vacuum fluctuations have a uniform spectral density,  $S_{n}^{\theta}(\Delta) = 1$, establishing a benchmark,
\begin{eqnarray}\label{eq:SNR_in}
\text{(SNR)}_{in,max} = 8\pi {|B_n|^2 \over \bar\Delta}.
\end{eqnarray}
%
%%%%%%%%%%%%%%%%%%%%%%%%%%%%%%%%%%%%%%
\subsubsection{Linear amplification}
Consider now the the linear amplification under nondegenerate resonance. The output signal here is, $C_n = u_n(0)B_n$, hence $\overline{P_0^\theta}= 8\pi G_n(0)|B_n|^2\cos^2(\theta-\theta_B - {\rm arg}\,u(0))$. 
The spectral density of the output noise is phase insensitive, 
\begin{eqnarray}\label{eq:S_ndeg}
S_n^\theta(\Delta) = (|u_n(\Delta)|^2 + |v_n(-\Delta)|^2) = G_n(\Delta) + G_m(\Delta) -1, \no
\end{eqnarray}
and gives the ratio, for large gain and sufficiently small bandwidth, $G_n(0)\gg1$,  $\bar\Delta \ll \delta_{th}$,
\begin{eqnarray}\label{eq:SNR_out_ndeg}
\text{(SNR)}_{out,max} \approx  8\pi {G_n(0)|B_n|^2 \over S_n^\theta(0)\bar \Delta} \approx  {1\over2} 
\text{(SNR)}_{in, max}.
\end{eqnarray}
This result reflects the fact that the signal and noise are equally amplified in the linear regime, and the reduction of the SNR by one half  is due to the noise contribution from the idler\cite{Caves1982,ClerkRMP2010}. 

For the degenerate resonance, one has to include the interference effect both for the signal and the noise. For the signal we have, $C_n(0) = u_n(0) B_n(0) + v_n(0) B^\ast_n(0)$, and  in the large-gain limit, 
\begin{eqnarray}\label{eq:P0_deg}
 \overline{P_0(\Delta)} &=& 32\pi G_n(0) |B_n(0)|^2 \cos^2 (\chi(0) - \theta) \no
&\times & \cos^2(\theta_B + \eta)\, ,
\end{eqnarray}
where $\chi(\Delta) = (1/2)\left({\rm arg}\,(u_n(\Delta)v_n(-\Delta)\right)$.
For the noise we get, using symmetries of the $uv$-coefficients, 
\begin{eqnarray}\label{eq:S_deg}
&&S_n^\theta(\Delta)   = e^{- 2r_n(\Delta)}
+\;  2\sinh 2r_n(\Delta) \cos^2\left(\chi(\Delta)- \theta\right) \,.  \no
\end{eqnarray}
The noise spectral density is illustrated in Fig.~\ref{Fig:Slinear} for the maximum amplification and squeezing directions.
For the large gain and small bandwidth this reduces to,
\begin{eqnarray}\label{eq:S_deg}
&&\overline{S_n^\theta(\Delta)}  
 = 4G_n(0) \cos^2\left(\chi(0) - \theta\right)\,\bar\Delta\,.
\end{eqnarray}
Here we see that the $\theta$-dependences of both signal and noise intensities are the same, giving relation,
\begin{eqnarray}\label{eq:SNR_out_deg}
\text{(SNR)}_{out,max} \approx  8\pi {|B_n|^2 \over\bar\Delta } =   \text{(SNR)}_{in, max}.
\end{eqnarray}
This result is consistent with the well-known property of linear amplification that there is no added noise under the phase sensitive amplification\cite{Caves1982,ClerkRMP2010}. 

\begin{figure}[tb]
\centering
\includegraphics[width=\columnwidth]{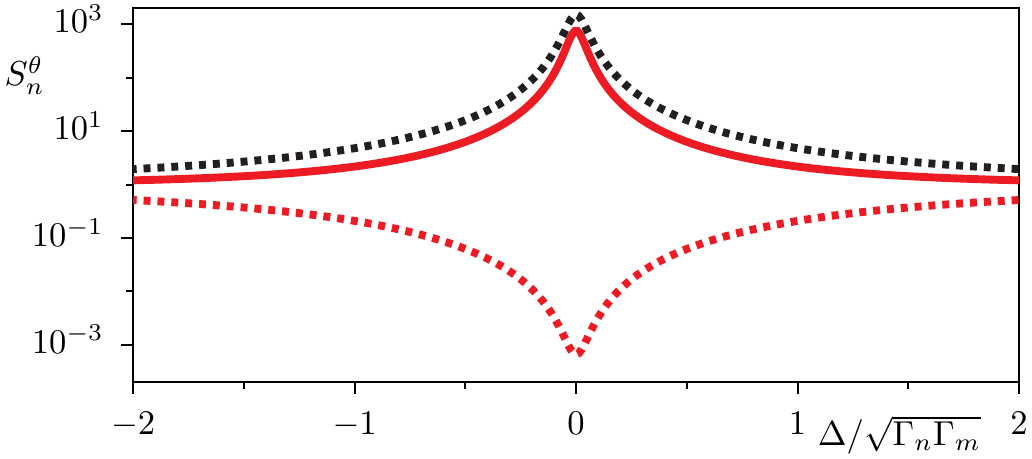}
\caption{Linear squeezing spectra $S_{n}^{\theta}(\Delta)$ vs. input detuning 
for two-mode squeezing under non-degenerate and degenerate resonance\cite{WustmannPRB2013,Wustmann2017}. Non-degenerate resonance 
(red solid);  degenerate resonance, $\theta=-\pi/4$ (black dashed) and  $\theta=\pi/4$ (red dashed). The amplification (squeezing) is localized in small detuning interval, $\Delta \ll \Gamma_j$ at large gain. 
[$\epsilon=0.95\sqrt{\Gamma_n\Gamma_m}$, $\delta=0$,  
$\Gamma_{m}=3\Gamma_{n}$,  $\Gamma_n=\Gamma_{n0}$.]
}
\label{Fig:Slinear}
\end{figure}
%

%%%%%%%%%%%%%%%%%%%%%%%%%%%%%%
\subsubsection{Nonlinear amplification}
In the nonlinear amplification regime the gains and optimal squeezing directions change for both the signal and the noise because of the Kerr effect. The magnitude of the Kerr frequency shifts are however different for signal and noise (compare  \Eq{zeta_ndeg}, with \Eqs{eq:EOM_A_linearized} and \eqref{eq:zetas_abovethresh}). This difference can be seen in Fig.~\ref{fig:Csq_vs_Bsq_deg}: because of a convex shape of the output vs input curve, the differential gain, which characterizes the noise, is smaller than the gain of a strong signal. Therefore, one should anticipate enhanced SNR value in the nonlinear regime.  This argument, however, does not take into account additional effect of the squeezing:  the anisotropy of the signal gain and of the noise spectral density do not generally coincide. Ideally one would wish that the direction of noise squeezing would be close to the direction of maximum signal amplification. Unfortunately, for the degenerate resonance it is not the case\cite{WustmannPRB2013}. A comparison of $uv$-coefficients for signal and noise at large gain close to the instability threshold, $\delta_{th} \ll \zeta_n \ll \Gamma_n$, shows that the $\theta$-anisotropy of both quantities is similar, see Fig.~\ref{fig:S_4mode}(a). Therefore SNR is given by the ratio of gains, for signal 
$G_n(0) =  (2\Gamma_n^2/\zeta_n^2)^2$, and for noise, $\tilde G_n(0) = (2\Gamma_n^2/3\zeta_n^2)^2$, thus
\begin{eqnarray}\label{eq:SNR_out_deg_nlin}
\text{(SNR)}_{out,max} \approx  72\pi {|B_n|^2 \over\bar\Delta } =  9 \text{(SNR)}_{in, max}.
\end{eqnarray}
This is nine times {\em larger} than the linear result, \Eq{eq:SNR_out_deg}. Numerical computation presented in Fig.~\ref{fig:S_4mode}(a) supports this analytical result. Here the bold lines, indicating the output squeezing power $S_n^{\theta}(\Delta=0)$ (red solid), and the normalized signal power $\overline{P_0}/|B_n|^2$  (dark green dashed) have rather close squeezing directions. In comparison, the thin lines show these quantities in the linear approximation, $\znl_n = \zeta_n = \delta$ and $\enl_n = \epsilon_n$, when the squeezing directions exactly coincide.
\begin{figure}[h]
\centering
  \includegraphics[width=\columnwidth]{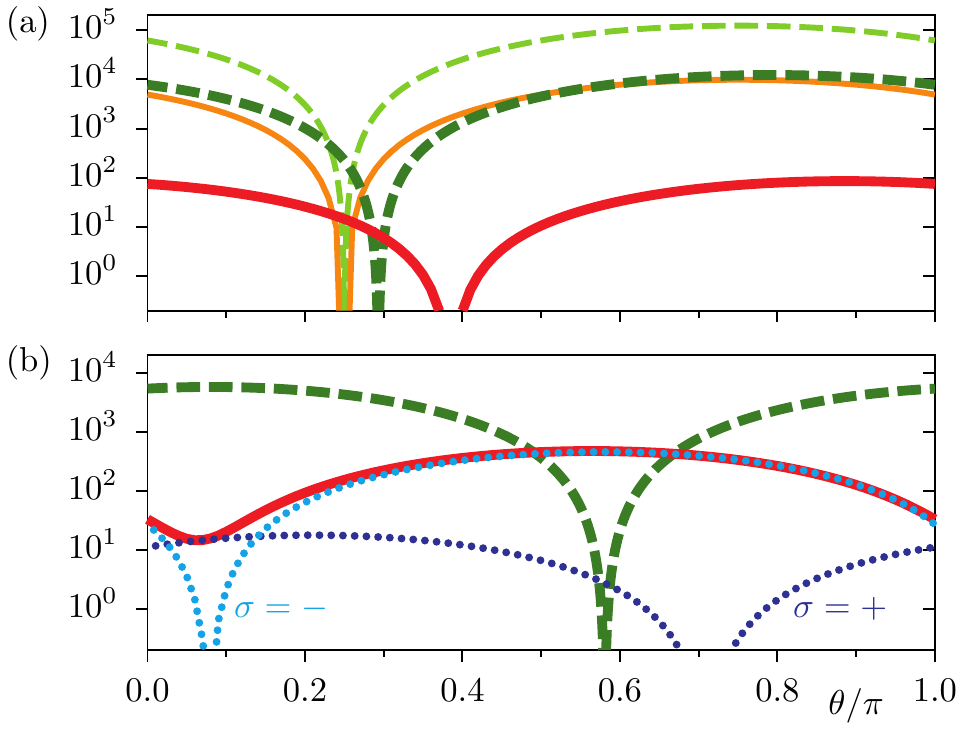}
\caption{Squeezing of output quantum noise spectral density and strongly amplified on-resonance signal, (a) for degenerate resonance at $\epsilon=0.95\Gamma$, and (b) for non-degenerate resonance at $\epsilon=\Gamma$\cite{WustmannPRB2013,Wustmann2017}. 
Solid red lines refer to quantum noise squeezing spectrum $S_{n}^\theta(0)$,  bold green dashed lines
refer to classical quadrature response $\overline{P_0^\theta}/|B_n|^2$.
In (a) the squeezing directions almost coinside for the signal and noise; thin lines show $S_{n}^\theta(0)$ (orange solid line) and $\overline{P_0}^\theta/|B_n|^2$ (light green dashed line) in the linear approximation. In (b) blue dotted lines show misoriented supermode contributions to $S_{n}^\theta(0)$; squeezing direction of noise almost coincides with the signal maximum amplification.
[$\delta=0,  |B_n|^2 = 0.1\Gamma, \theta_B = 0, 
\Gamma_n=\Gamma_m,
\alpha_n = \alpha_m = \Gamma/100$, $\Gamma_n=\Gamma_{n0}$.]
}
\label{fig:S_4mode}
\end{figure}

 For the non-degenerate resonance the analysis is more complex due to four-mode squeezing, it can only be done numerically  or analytically for the balanced mode model\cite{Wustmann2017}. The result of numerics is shown in Fig.~\ref{fig:S_4mode}(b)
for the representative case of signal input, $|B_n|^2 = 0.1\Gamma_n$, at the threshold, $\epsilon_n=\Gamma_n$. 
The figure compares $S_{n}^\theta( 0)$ (red line) with the relative spectral power of the signal $\overline{P_0^\theta}/|B_n|^2$ (green dashed line). The contributions from supermodes have different squeezing directions, shifted by 
more than $\pi/2$. While the squeezing direction of  the $\sigma=+$ supermode (dark blue dotted line) is close to the squeezing direction of the signal, the squeezing direction of the dominant $\sigma=-$ supermode (light blue dotted line) approximately coincides with the maximum of signal amplification. This results in strong suppression of the overall noise in the direction of the maximum signal amplification. The maximum SNR value is achieved at $\theta=\theta_B+ 0.06\pi$, where $\overline P_0 \approx  1820\pi |B_n|^2$ and $S_{nn}(0) \approx 14.5$, giving,
\begin{eqnarray}\label{}
(\text{SNR})_{out,max} &\approx& 125 \pi \,{|B_n(0)|^2\over \bar\Delta} \,. 
\end{eqnarray}
This is about 30 times larger than the linear result, \Eq{eq:SNR_out_ndeg}, and 15 times larger than the input value.
 
%%%%%%%%%%%%%%%%%%%%%%%%%%%
\section{Concluding comments}
Parametric effects in c-QED is a rich and interesting field of research with great potential for applications in quantum information technology.  The field is far from being fully explored. Some phenomena, such as quantum limited  parametric amplification and frequency conversion are already included in c-QED toolbox. Other phenomena are waiting for their exploration, for example,  phase locking and synchronization effects under non-degenerate parametric resonance and subharmonic oscillations, or experimental testing of the effect of enhancement of SNR under nonlinear amplification. 

In this review we restricted to the classical description of nonlinear parametric effects and small quantum fluctuations. Some interesting results on  large fluctuations reported in literature are left outside our discussion.   
This concerns  critical fluctuations in the vicinity of parametric threshold\cite{KryKhe1996,Drummond2002,MilburnDuty2014},  and transitions between  phase degenerate oscillatory states\cite{MarthalerPRA2006,MarthalerPRA2007,DykmanPRL2012,AnkerholdPRA2017}. 
Some problems related to large quantum fluctuations are not yet explored, for instance quantum statistics of nondegenerate parametric oscillation, or dissipative phase transition in the bistability regime of subharmonic oscillations.
Another interesting direction to explore is the possibility to use  degenerate states of Josephson parametric oscillators coupled by mutual injection locking for quantum simulation as it is done in a quantum optics\cite{YamamotoPRA2015}.

Parametric effects in c-QED is an excellent playground for testing the possibility of quantum information processing with  
continuous variables. By encoding quantum information in the oscillator bosonic states  rather than discrete states of conventional qubits, one can envision efficient computational protocols and error correction schemes\cite{GottesmanPRA2001}. Recently proposed implementation of these ideas with c-QED\cite{MirrahimiNJP2014}  is closely connected to the physics of parametric oscillatory states discussed here. Generation and stabilization of quantum cat states under degenerate parametric resonance was recently studied\cite{Puri2017,CiutiPRA2016}, similar questions can be addressed  regarding multicomponent subharmonic oscillatory states.

The scope of this review is limited to the c-QED field. However, many related quantum parametric effects are also available in quantum acoustics and  optomechanics. For instance, the non-degenerate parametric resonance in a hybrid optomechanical resonator may involve one mechanical and one electromagnetic degree of freedom\cite{ShumeikoPRA2016}. In fact the quantum amplification and frequency conversion are observed in such devices in the microwave domain\cite{SillampaaNature2013,PalomakiScience2013,LecocqPRL2016,SillampaaPRX2016}. The same non-degenerate parametric oscillation effects as described in Sec.~\ref{sec:ndeg_osc} were observed in a mechanical oscillator   \cite{SunNatCom2016}. Furthermore, the  demonstration of the  frequency conversion between the microwaves and  telecom optics in parametrically driven optomechanical resonators\cite{ClelandNatPhys2013,LehnertNatPhys2014}  paves a way for integrating c-QED  devices in a long distance quantum communication network. 

\section{Acknowledgement}
We are thankful to many people for the discussions we enjoyed during various stages of this work, in particular, to Per Delsing, Chris Wilson, Mark Dykman, Andreas Bengtsson, and Ida-Maria Svensson. The support from the Knut and Alice Wallenberg Foundation is gratefully acknowledged.

%%%%%%%%%%%%%%%%%%%%%%%%%%%%%%%


\begin{thebibliography}{}


%%%%%%%%%%%%%%%%%%%%%%
%%%     INTRODUCTION

\bibitem{ChowNatC2014}
J.M. Chow, % {\em et al.}, % Implementing a strand of a scalable fault-tolerant quantum computing fabric 
J.M.~Gambetta, E.~Magesan, D.W.~Abraham, A.W.~Cross, B.R.~Johnson, N.A.~Masluk, C.A.~Ryan, J.A.~Smolin, S.J.~Srinivasan, and  M.~ Steffen,
Nat. Commun. {\bf 5}, 4015 (2014).

\bibitem{DiCarloNatC2017}
N. K. Langford, %{\em et al.},
R. Sagastizabal, M. Kounalakis, C. Dickel,A. Bruno, F. Luthi, D. J. Thoen, A. Endo, and L. DiCarlo
Experimentally simulating the dynamics of quantum light and matter at ultrastrong coupling
Nat. Comm. {\bf 8}, 1715 (2017). 

%Revews

\bibitem{MakhlinRMP2001}
Yu. Makhlin, G. Sch\"on, and A. Shnirman %Quantum-state engineering with Josephson-junction devices
Rev. Mod. Phys. {\bf73}, 357 (2001).
 
 \bibitem{MartinisArxiv2004}
M.H. Devoret, A. Wallraff, and J.M. Martinis, 
 {\em Superconducting qubits: a short review},
arXiv:cond-mat/0411174 ( 2004).
 
 \bibitem{WendinLTP2007}
G. Wendin and V.S. Shumeiko, Low Temp. Phys, {\bf 33}, 724 (2007).

\bibitem{SchoelkopfNat2008} %wiring
R.J.~Schoelkopf and S.M.~Girvin, Nature {\bf 451}, 664 (2008).

\bibitem{KockumPR2017}
X. Gu, A. F. Kockum, A. Miranowicz, Y.-X. Liu, and F. Nori, %Microwave photonics with superconducting quantum circuits 
Phys. Rep. {\bf 718–719}, 1 (2017).

\bibitem{WendinRPP2017}
G. Wendin, %Quantum information processing with superconducting circuits: a review
Rep. Prog. Phys. {\bf80}, 106001 (2017). 

%Amplifiers

\bibitem{BergealNature2010}
N.~Bergeal, %{\em et al.,}
F.~Schackert, M.~Metcalfe, R.~Vijay, V.E.~Manucharyan, L.~Frunzio, D.E.~Prober, R.J.~Schoelkopf, S.M.~Girvin, and M.H.~Devoret,
Nature {\bf 465}, 64 (2010).


%2mode squeezing
\bibitem{EichlerPRL2011}
C.~Eichler,  %{\em et al.,}
D.~Bozyigit, C.~Lang, M.~Baur, L.~Steffen, J.M.~Fink, S.~Filipp, and A.~Wallraff,
Phys. Rev. Lett. {\bf 107}, 113601 (2011).

%Entanglement
\bibitem{BergealPRL2012}%2mode correlation
N.~Bergeal, F.~Schackert, L.~Frunzio, and M.H.~Devoret,
Phys. Rev. Lett. {\bf 108}, 123902 (2012).

\bibitem{FluETAL2012}%entangl in 2 TL
E.~Flurin, N.~Roch, F.~Mallet, M.H.~Devoret, and B.~Huard,
Phys. Rev. Lett. {\bf 109}, 183901 (2012).

\bibitem{RocETAL2012}%2mode correlation
N.~Roch, %{\em et al.,}
E.~Flurin, F.~Nguyen, P.~Morfin, P.~Campagne-Ibarcq, M.H.~Devoret, and B.~Huard,
Phys. Rev. Lett. {\bf 108}, 147701 (2012).

\bibitem{EichlerPRL2012}%entangl itinerant
C.~Eichler, %{\em et al.,}
C.~Lang, J.M.~Fink, J.~Govenius, S.~Filipp, and A.~Wallraff,
Phys. Rev. Lett. {\bf 109}, 240501 (2012).

%Flux JPA
\bibitem{YamETAL2008}%JPA tunable cav
T.~Yamamoto, %{\em et al.,}
K.~Inomata, M.~Watanabe, K.~Matsuba, T.~Miyazaki, W.D.~Oliver, Y.~Nakamura, and J.S.~Tsai,
Appl. Phys. Lett. {\bf 93}, 042510 (2008).

\bibitem{VijayPRL2011} %Observation of Quantum Jumps in a Superconducting Artificial Atom
R. Vijay, D.H. Slichter, and I. Siddiqi,
Phys. Rev. Lett. {\bf 106}, 110502 (2011)

\bibitem{RistePRL2012} % by Measurement of a Superconducting Quantum Bit Circuit
D. Riste, J. G. van Leeuwen, H.-S. Ku, K.W. Lehnert, and L. DiCarlo,
 Phys. Rev. Lett. {\bf109}, 050507 (2012).

\bibitem{MenzelPRL2012} %Path Entanglement of Continuous-Variable Quantum Microwaves
E. P. Menzel,  %{\em et al.,}
R. Di Candia, F. Deppe, P. Eder, L. Zhong, M. Ihmig, M. Haeberlein, A. Baust, E. Hoffmann, 
D. Ballester, K. Inomata, T. Yamamoto, Y. Nakamura, E. Solano, A. Marx, and R. Gross,
Phys. Rev. Lett. {\bf 109}, 250502 (2012).

\bibitem{NakamuraAPL2013}%Single-shot readout of  flux qubit with a flux-driven JPA
Z.R.~Lin, %{\em et al.,} 
K.~Inomata, W.D.~Oliver, K.~Koshino, Y.~Nakamura, J.S.~Tsai, and T.~Yamamoto,
Appl. Phys. Lett. {\bf 103}, 132602 (2013).

\bibitem{VionPRB2014}
X.~Zhou, %{\em et al.,}
V.~Schmitt, P.~Bertet, D.~Vion, W.~Wustmann, V.~Shumeiko, and D.~Esteve,
Phys. Rev. B {\bf 89}, 214517 (2014).

\bibitem{DeppeNJP2015} %Squeezing with a flux-driven Josephson parametric amplifier
L. Zhong, % {\em et al.,} 
E.P. Menzel, R.Di Candia, P. Eder, M. Ihmig, A. Baust, M. Haeberlein, E. Hoffmann, K. Inomata, T. Yamamoto, Y. Nakamura, E. Solano, F Deppe, A Marx, and R Gross,
New J. Phys {\bf 15},  125013 (2013).

%%%%%%%%%%%%%%%%%%%%%%%%%%%%%%%%%

\bibitem{GrossNatP2010}
 T. Niemczyk, %{\em et al.,} %Circuit quantum electrodynamics in the ultrastrong-coupling regime
 F. Deppe, H. Huebl, E. P. Menzel, F. Hocke, M. J. Schwarz, J. J. Garcia-Ripoll, D. Zueco, T. Hümmer, E. Solano, A. Marx, and R. Gross,
Nat. Phys. {\bf6}, 772 (2010).

\bibitem{LupascuNatP2017}%Ultrastrong coupling of a single artificial atom to an electromagnetic continuum in the nonperturbative regime
P. Forn-Díaz, %{\em et a.,} 
J. J. García-Ripoll, B. Peropadre, M. A. Yurtalan, J.-L. Orgiazzi, R. Belyansky, C. M. Wilson, and A. Lupascu, 
Nat. Phys. {\bf 13}, 39 (2017).
	
\bibitem{SembaNatP2017}
F. Yoshihara, T. Fuse, S. Ashhab, K. Kakuyanagi, Sh. Saito, and K. Semba,  %Superconducting qubit-oscillator circuit beyond the ultrastrong-coupling regime
Nat. Phys. {\bf 13}, 44 (2017).	

\bibitem{WilsonNature2011} %DCE
C.M.~Wilson, %{\em et al.,}
G.~Johansson, A.~Pourkabirian, M.~Simoen, J.R.~Johansson, T.~Duty, F.~Nori, and P.~Delsing,
Nature {\bf 479}, 376 (2011).

\bibitem{VistakisSci2013} %cats
B. Vlastakis {\em et al.,} Science {\bf 342}, 607 (2013).

\bibitem{LeghtasSci2015}
Z. Leghtas, %{\em et al.,} % Confining the state of light to a quantum manifold by engineeredtwo-photon loss
S. Touzard, I.M. Pop, A. Kou, B. Vlastakis, A. Petrenko, K.M. Sliwa, A. Narla, S. Shankar, M.J. Hatridge, M. Reagor, L. Frunzio, R.J. Schoelkopf, M. Mirrahimi, and M.H. Devoret,	
Science {\bf 347}, 853 (2015).

\bibitem{Wallquist2006}%Tunable cavity
M.~Wallquist, V.S.~Shumeiko, and G.~Wendin,
Phys. Rev. B {\bf 74}, 224506 (2006).

\bibitem{Sandberg2008} %Tunable cavity
M.~Sandberg, %{\em et al.,}
C.M.~Wilson, F.~Persson, T.~Bauch, G.~Johansson, V.~Shumeiko, T.~Duty, and P.~Delsing,
Appl. Phys. Lett. {\bf 92}, 203501 (2008).

\bibitem{WilsonPRL2010}
C.M.~Wilson, %{\em et al.,}
 T.~Duty, M.~Sandberg, F.~Persson, V.~Shumeiko, and P.~Delsing,
Phys. Rev. Lett. {\bf 105}, 233907 (2010).

\bibitem{AspelmeyerRMP2014}
M. Aspelmeyer, T.J. Kippenberg, and F. Marquardt, %Cavity optomechanics
Rev. Mod. Phys. {\bf 86}, 1391 (2014).

\bibitem{Moore1970}%DCE
G.~T.~Moore, J. Math. Phys. {\bf 11}, 2679 (1970).

 \bibitem{JohanssonPRL2009}%DCE
 J.R.~Johansson, G.~Johansson, C.M.~Wilson, and F.~Nori,
 PRL {\bf 103}, 147003 (2009).

\bibitem{JohanssonPRA2010}%DCE
J.R.~Johansson, G.~Johansson, C.M.~Wilson, and F.~Nori,
PRA {\bf 82}, 052509 (2010).

\bibitem{NayfehBook}
A.H.~Nayfeh and D.T.~Mook, {\em Nonlinear Oscillations}
(Wiley, New York, 1979).

\bibitem{BogoliubovBook}
N.N. Bogoliubov and Y.A. Mitropolsky,  {\em Asymptotic Methods in the Theory of Non-Linear Oscillations} (Gordon \& Breach, 1961).

\bibitem{JordanBook}
D. W. Jordan and P. Smith, {\em Nonlinear Ordinary Differential
Equations} (Oxford University Press, Oxford, 2007).

\bibitem{AndreasLic2017}
A.Bengtsson, {\it Degenerate and nondegenerate Josephson parametric oscillators for quantum information applications},
Licentiate Thesis, Chalmers University of Technology (2017). 

\bibitem{WustmannPRB2013}
W.~Wustmann and V.~Shumeiko, 
Phys. Rev. B {\bf 87}, 184501 (2013).

\bibitem{YurkeDen1984}
B.~Yurke and J.S.~Denker, Phys. Rev. A {\bf 29}, 1419 (1984).
  
\bibitem{Devoret2004}
M.H.~Devoret, in {\em Quantum Entanglement and Information Processing},
edited by D.~Esteve, J.M.~Raimond, and J.~Dalibard, 
Proceedings of the Les Houches Summer School of Theoretical Physics, 
LXIII, 1995 (Elsevier, Amsterdam, 2004).

\bibitem{HasslerPRB2016}
J.~Ulrich and F.~Hassler %Dual approach to circuit quantization using loop charges
Phys. Rev. B {\bf 94}, 094505 (2016).

\bibitem{GarZol2000}
 C.~W.~Gardiner and P.~Zoller,
 {\em Quantum Noise}, (Springer, Berlin, 2000).
 
 \bibitem{ColGar1984}
M.J.~Collett and C.W.~Gardiner, 
Phys. Rev. A {\bf 30}, 1386 (1984).

\bibitem{Yurke_DruFic2004}
B.~Yurke %: {\em Input-Output Theory} 
in {\em Quantum Squeezing}, 
edited by P.D.~Drummond and Z.~Fizek (Springer, Berlin, 2004).

\bibitem{Wustmann2017}
W. Wustmann and V. Shumeiko, Phys. Rev. Appl. {\bf 8}, 024018 (2017). 
 
\bibitem{DykmanPRE1998} %Fluctuational phase-flip transitions in parametrically driven oscillators
M.I.~Dykman, C.M.~Maloney, V.N.~Smelyanskiy, and M.~Silverstein,
Phys. Rev. E {\bf 57}, 5202 (1998).


\bibitem{CasETAL2008}% Paramp with array 480 SQUIDS
M.A.~Castellanos-Beltran, K.D.~Irwin, G.C.~Hilton, L.R.~Vale and K.W.~Lehnert,
Nature Phys. {\bf 4}, 929 (2008).

\bibitem{HakonenPNAS2013}
%Dynamical Casimir effect in a Josephson metamaterial
P. Lähteenmäki, G.S. Paraoanu, J. Hassel, and P.J. Hakonen,
PNAS {\bf 110}, 4234 (2013). 

\bibitem{SvenssonLT2018}%Microwave photon generation in a doubly tunable superconducting resonator
I-M. Svensson, % {et al.,}, 
M. Pierre, M. Simoen, W. Wustmann, P. Krantz, A. Bengtsson, G. Johansson, J. Bylander, V. Shumeiko and P. Delsing, 
J. of Phys: Conf. Ser. {\bf 969}, 012146 (2018). 

\bibitem{SvenssonPhD}
I.-M. Svensson, {\em Tunable Superconducting Resonators}, PhD Thesis, Chalmers University of Technology, (2018).

\bibitem{BengtssonMS2015}
A. Bengtsson, {\em Parametric frequency conversion in two coupled superconducting resonators}, Master Thesis, Chalmers University of Technology, (2015).

\bibitem{LambrechtPRL1996}%DCE
A. Lambrecht, M.-T. Jaekel, and S. Reynaud,
Phys. Rev. Lett. {\bf 77}, 615 1996)

\bibitem{YurkePRA1989} %Observation of parametric amplification and deamplification in a Josephson parametric amplifier
B. Yurke,  %{\em et al.,}
L.R. Corruccini, P.G. Kaminsky, and L.W. Rupp, A.D. Smith, A.H. Silver, and R.W. Simon, and E.A. Whittaker,
Phys. Rev. A {\bf 39}, 2519 (1989).

%\bibitem{LinNatCom2014} %Tuneble Paramp +locking
%Z.R. Lin,  %{\em et al.,}
%K. Inomata, K. Koshino, W.D. Oliver, Y. Nakamura, J.S. Tsai, and T. Yamamoto, 
%Nat. Commun. {\bf 5}, 4480 (2014).

\bibitem{RoyAmplification}%Introduction to Quantum-limited Parametric Amplification of Quantum Signals with Josephson Circuits
M. Devoret and  A. Roy, 
Comptes Rendus Physique, {\bf 17}, 740 (2016); arXiv[cond.-mat]:1605.00539. 


\bibitem{Caves1982}
C.M.~Caves, Phys. Rev. D {\bf 26}, 1817 (1982).

\bibitem{ClerkRMP2010}
A.A.~Clerk, M.H.~Devoret, S.M.~Girvin, F.~Marquardt, and R.J.~Schoelkopf, 
Rev. Mod. Phys. {\bf 82}, 1155 (2010).

%Multimode 
\bibitem{BraunsteinPRA2005}
S.L.~Braunstein,  Phys. Rev. A {\bf 71}, 055801 (2005).
 
\bibitem{TrepsD2010}
G.~Patera, N.~Treps, C.~Fabre, and G.J.~de Valcarcel, 
Eur. Phys. J. D {\bf 56}, 123 (2010).

\bibitem{FabreJdP1989}
C.~Fabre, E.~Giacobino, A.~Heidmann, and S.~Reynaud,  
J. de Physique, {\bf 50}, 1209 (1989).
 
\bibitem{AumentadoNaPh2011}%conversion cavity
E.~Zakka-Bajjani, F.~Nguyen, M.~Lee, L.R.~Vale, R.W.~Simmonds, and J.~Aumentado, 
Nature Phys.  {\bf 7}, 599 (2011).

\bibitem{AbdoPRL2013}%conversion propagating
B.~Abdo, %{\em et al.,}
 K.~Sliwa, F.~Schackert, N.~Bergeal, M.~Hatridge, L.~Frunzio, A.D.~Stone, and M.~Devoret, 
Phys. Rev. Lett. {\bf 110}, 173902 (2013).

\bibitem{AumentadoAPL2015} %Coherent-state storage and retrieval between superconducting cavities using parametric frequency conversion
A.J. Sirois, %{\em et al.,} 
M.A. Castellanos-Beltran, M.P. DeFeo, L. Ranzani, F. Lecocq, R.W. Simmonds, J.D. Teufel, and J. Aumentado,
Appl. Phys. Lett. {\bf 106}, 172603 (2015).

\bibitem{SvenssonPRB2017} %Period-tripling subharmonic oscillations in a driven superconducting resonator,”
I.-M. Svensson, A. Bengtsson, P. Krantz, J. Bylander, V. Shumeiko, and P. Delsing,  
Phys. Rev. B {\bf 96}, 174503 (2017).

\bibitem{Landau}
L.D. Landau and E.M. Lifshits, {\em Mechanics} (Butterworth-Heinemann Ltd., 1982).

\bibitem{Haken} %Cooperative phenomena far from equilibrium
H. Haken, Rev. Mod. Phys. {\bf 47} 67 (1975).

\bibitem{BengtssonPRB2018} %Nondegenerate parametric oscillations in a tunable superconducting resonator
A. Bengtsson,  %{\em et al.,}
 Ph. Krantz, M. Simoen, I.-M. Svensson, B. Schneider, V. Shumeiko, P. Delsing, and J. Bylander,
Phys. Rev. B {\bf 97}, 144502 (2018).

\bibitem{KryKhe1996}
G.Yu.~Kryuchkyan and K.V.~Kheruntsyan,
Optics Communications {\bf 127}, 230 (1996).

\bibitem{Drummond2002}
P. D. Drummond, K. Dechoum, and S. Chaturvedi, Phys. Rev. A {\bf 65}, 033806 (2002).

\bibitem{MilburnDuty2014} %Quantum and classical nonlinear dynamics in a microwave cavity
C.H.~Meaney, H.~Nha, T.~Duty, and G.J.~Milburn,
EPJ Quantum Techn. {\bf 1}(7) (2014)

\bibitem{MarthalerPRA2006} 
%Switching via quantum activation: A parametrically modulated oscillator
M.~Marthaler and M.I.~Dykman,
Phys. Rev. A {\bf 73}, 042108 (2006).

\bibitem{MarthalerPRA2007} 
%Quantum interference in the classically forbidden region: A parametric oscillator
M.~Marthaler and M.I.~Dykman,
Phys. Rev. A {\bf 76}, 010102 (2007).

\bibitem{DykmanPRL2012} 
%Sharp Tunneling Peaks in a Parametric Oscillator: Quantum Resonances Missingin the RotatingWave Approximation
V.~Peano, M.~Marthaler, and M.I.~Dykman
Phys. Rev. Lett. {\bf 109}, 090401 (2012).

\bibitem{Cochrane1999}
P.T.~Cochrane, G.J.~Milburn, and W.J.~Munro, Phys. Rev. A {\bf 59}, 2631 (1999).

\bibitem{Puri2017}
S. Puri, S. Boutin, and A. Blais, npj Quantum Inf. 3, 18 (2017).

\bibitem{LinNatComm2014} % Locking JPA
Z.R. Lin, %{\em et al.,}
K. Inomata, K. Koshino, W. D. Oliver, Y. Nakamura, J.-S. Tsai, and T. Yamamoto, 
Nat. Commun. {\bf 5}, 4480 (2014).

\bibitem{KrantzNatComm2016}
Ph.~Krantz, %{\em et al.,}
A.~Bengtsson, M.~Simoen, S.~Gustavsson, V.~Shumeiko, W.D.~Oliver, C.M.~Wilson, P.~Delsing, and J.~Bylander, 
Nature Communication  {\bf 7}, 11417 (2016). 

\bibitem{RytovBook}
S.M.~Rytov, Yu.A.~Kravtsov, and V.I.~Tatarskii, 
{\em Principles of Statistical Radiophysics 1} (Springer-Verlag Berlin, 2011). 

\bibitem{ReynaudJMO1991}
J.L.~Courtois, A.~Smith, C.~Fabre, and S.~Reynaud,
J. Mod. Opt. {\bf 38}, 177 (1991).

\bibitem{Adler1946}
R. Adler, Proc. IRE {\bf 34}, 351 (1946).

\bibitem{Hayashi}
C. Hayashi, {\em Nonlinear Oscillations in Physical Systems} (Princeton
University Press, Princeton, NJ, 1985).

 \bibitem{GuoPRL2013}
 L. Guo, M. Marthaler, and G. Sch\"on, 
 Phys. Rev. Lett. {\bf111}, 205303 (2013). 

\bibitem{SvenssonAPL2018} %Period multiplication in a parametrically driven superconducting resonator
I.-M.~Svensson, A.~Bengtsson, J.~Bylander, V.~Shumeiko, and P.~Delsing,
Appl. Phys. Lett. {\bf 113}, 022602 (2018).

\bibitem{SvenssonLicense}
Reproduced from \cite{SvenssonPRB2017} %I.-M.~Svensson, {et al.,}, Appl. Phys. Lett. {\bf 113}, 022602 (2018), 
with the permission of AIP Publishing.

\bibitem{SvenssonPRB2017} %Period-tripling subharmonic oscillations in a driven superconducting resonator
I.-M. Svensson, A.~Bengtsson, Ph.~Krantz, J.~Bylander, V.~Shumeiko, and P.~Delsing,
Phys. Rev. B {\bf 96}, 174503 (2017).

\bibitem{SqueezingBook}
{\em Quantum Squeezing}, 
edited by P.D.~Drummond and Z.~Fizek (Springer, Berlin, 2004).

%Disentaglement
\bibitem{Perelomov1977}
A.M.~Perelomov, Sov. Phys. Usp. {\bf 20} 703 (1977).
 
 \bibitem{Collet1988}
M.J.~Collett, Phys. Rev. A {\bf 38}, 2233 (1988).

\bibitem{AnkerholdPRA2017} %Time-translation-symmetry breaking in a driven oscillator: From the quantum coherent to the incoherent regime, 
Y. Zhang, J. Gosner, S.M. Girvin, J. Ankerhold, and M.I. Dykman,
Phys. Rev. A {\bf 96}, 052124 (2017).

\bibitem{SaraivaPRL2017} %Photonic Counterparts of Cooper Pairs
A. Saraiva, %{\em et al.,}
F.S. de Aguiar Júnior, R. de Melo e Souza, and A. Patrocínio Pena,
Phys. Rev. Lett. {\bf 119}, 193603 (2017).

\bibitem{BCS}
J.~Bardeen, L.N.~Cooper, and J.R.~ Schrieffer,  Phys. Rev. {\bf 108} 1175 (1957). 

\bibitem{Entropy}
S.J. van Enk, Phys. Rev. A {\bf 60}, 5095 (1999).

%%%%%%%%%%%%%%
%Summary


\bibitem{YamamotoPRA2015}%Quantum correlation in degenerate optical parametric oscillators with mutual injections
K.~Takata, A.~Marandi, and Y.~Yamamoto,
Phys. Rev. A {\bf 92}, 043821 (2015).

\bibitem{GottesmanPRA2001} %Encoding a qubit in an oscillator
D.~Gottesman, A.~Kitaev, and J.~ Preskill,
Phys. Rev. A {\bf 64}, 012310 (2001).

\bibitem{MirrahimiNJP2014}
M. Mirrahimi, {\em et al.,}
Z. Leghtas, V.V. Albert, S. Touzard, R.J. Schoelkopf, L. Jiang, and M.H. Devoret, 
New J. Phys. {\bf16}, 045014 (2014).

\bibitem{CiutiPRA2016} %Exact steady state of a Kerr resonator with one- and two-photon driving and dissipation:Controllable Wigner-function multimodality and dissipative phase transitions
N.~Bartolo, F.~Minganti, W.~Casteels, and C.~Ciuti,
Phys. Rev. A {\bf 94}, 033841 (2016).

\bibitem{ShumeikoPRA2016}
V.~Shumeiko, Phys. Rev. A {\bf 93}, 023838 (2016).

%OM
\bibitem{SillampaaNature2013}% Hybrid circuit cavity quantumelectrodynamics with a micromechanical resonator
J.-M. Pirkkalainen, S.U. Cho, Jian Li, G.S. Paraoanu, P. J. Hakonen, and M.A. Sillanp\"a\"a,
Nature {\bf494}, 211 (2013).

\bibitem{PalomakiScience2013}% “Entangling mechanical motion with microwave fields,”
T. A. Palomaki, J. D. Teufel, R. W. Simmonds, and K. W. Lehnert,  
Science {\bf 342}, 710 (2013).

\bibitem{LecocqPRL2016} %“Mechanically mediated microwave frequency conversion in the quantum regime,”
F. Lecocq, J.B. Clark, R.W. Simmonds, J. Aumentado, and J.D. Teufel, 
Phys. Rev. Lett. {\bf 116}, 043601 (2016).

\bibitem{SillampaaPRX2016} %“Low-noise amplification and frequency conversion with a multiport microwave optomechanical device,” 
C.F. Ockeloen-Korppi, %{\em et al,.}
E. Damsk\"agg, J.M. Pirkkalainen, T.T. Heikkil\"a, F. Massel, and M.A. Sillanp\"a\"a,  
Phys. Rev. X {\bf 6}, 041024 (2016).

\bibitem{SunNatCom2016}%OM nondegenerate oscillation
F. Sun, X. Dong, J. Zou, M. I. Dykman, and H. B. Chan, 
Nat. Commun. {\bf 7}, 12694 (2016).

\bibitem{ClelandNatPhys2013}%Nanomechanical coupling between microwave and optical photons
J. Bochmann, A. Vainsencher, D.D. Awschalom, and A.N. Cleland, 
Nat. Phys. {\bf 9}, 712 (2013).

\bibitem{LehnertNatPhys2014}%Bidirectional and ecient conversion between microwave and optical light
R.W. Andrews, %{\em et al.,}, 
R.W. Peterson, T.P. Purdy, K. Cicak, R.W. Simmonds, C.A. Regal and K.W. Lehnert,
Nat. Phys. {\bf 10}, 321 (2014).



%%%%%%%%%%%%%%%%%%%%%%%%%%%%%%%%

\end{thebibliography}
\end{document}